\documentclass[fleqn,usenatbib]{mnras}

\usepackage{newtxtext,newtxmath}

\usepackage[T1]{fontenc}

\DeclareRobustCommand{\VAN}[3]{#2}
\let\VANthebibliography\thebibliography
\def\thebibliography{\DeclareRobustCommand{\VAN}[3]{##3}\VANthebibliography}

\usepackage{graphicx}	
\usepackage{amsmath}	
\usepackage{float}
\usepackage{url}

\defcitealias{Mertens2020}{M20}
\graphicspath{{./}{figures/}}
\usepackage{xcolor}

\newcommand{\revised}[1]{{ #1}}

\title[Testing ML+GPR]{21-cm Signal from the Epoch of Reionization: A Machine Learning upgrade to Foreground Removal with Gaussian Process Regression}

\author[A. Acharya et al.]{Anshuman Acharya$^{1}$\thanks{E-mail: anshuman@mpa-garching.mpg.de},
Florent Mertens$^{2}$,
Benedetta Ciardi$^{1}$,
Raghunath Ghara$^{3}$,       
\newauthor L\'eon V. E. Koopmans$^{4}$, Sambit K. Giri$^{5}$, Ian Hothi$^{2,6}$, Qing-Bo Ma$^{7,8}$, Garrelt Mellema$^{9}$, \newauthor Satyapan Munshi$^{4}$
\\
$^{1}$Max-Planck-Institut für Astrophysik, Garching 85748, Germany \\
$^{2}$LERMA, Observatoire de Paris, PSL Research University, CNRS, Sorbonne Universit\'{e}, F-75014 Paris, France \\
$^{3}$Astrophysics Research Centre, Open University of Israel, Ra’anana 4353701, Israel \\
$^{4}$Kapteyn Astronomical Institute, University of Groningen, P.O. Box 800, 9700AV Groningen, The Netherlands \\
$^{5}$Nordita, KTH Royal Institute of Technology and Stockholm University, Hannes Alfv\'ens v\"ag 12, SE-106 91 Stockholm, Sweden \\
$^{6}$Laboratoire de Physique de l’ENS, ENS, Universit\'{e} PSL, CNRS, Sorbonne Universit\'{e}, Universit\'{e}e Paris Cit\'{e}, 75005, Paris, France \\
$^{7}$School of Physics and Electronic Science, Guizhou Normal University, Guiyang 550001, PR China \\
$^{8}$Guizhou Provincial Key Laboratory of Radio Astronomy and Data Processing, Guizhou Normal University, Guiyang 550001, PR China \\
$^{9}$The Oskar Klein Centre, Department of Astronomy, Stockholm University, AlbaNova, SE-10691 Stockholm, Sweden
}

\date{Accepted XXX. Received YYY; in original form ZZZ. NORDITA-2023-074}

\pubyear{2023}

\begin{document}
\label{firstpage}
\pagerange{\pageref{firstpage}--\pageref{lastpage}}
\maketitle

\begin{abstract}
In recent years, a Gaussian Process Regression (GPR) based framework has been developed for foreground mitigation from data collected by the LOw-Frequency ARray (LOFAR), to measure the 21-cm signal power spectrum from the Epoch of Reionization (EoR) and Cosmic Dawn. However, it has been noted that through this method there can be a significant amount of signal loss if the EoR signal covariance is misestimated. To obtain better covariance models, we propose to use a kernel trained on the {\tt GRIZZLY} simulations using a Variational Auto-Encoder (VAE) based algorithm. In this work, we explore the abilities of this Machine Learning based kernel (VAE kernel) used with GPR, by testing it on mock signals from a variety of simulations, exploring noise levels corresponding to $\approx$10 nights ($\approx$141 hours) and $\approx$100 nights ($\approx$1410 hours) of observations with LOFAR. Our work suggests the possibility of successful extraction of the 21-cm signal within 2$\sigma$ uncertainty in most cases using the VAE kernel, with better recovery of both shape and power than with previously used covariance models. We also explore the role of the excess noise component identified in past applications of GPR and additionally analyse the possibility of redshift dependence on the performance of the VAE kernel. The latter allows us to prepare for future LOFAR observations at a range of redshifts, as well as compare with results from other telescopes.
\end{abstract}



\section{Introduction}
The Epoch of Reionization (EoR) follows the period of Cosmic Dawn, when the first stars, galaxies, black holes and other astrophysical objects formed. These objects began to radiate photons that ionized the neutral gas across the Universe. Understanding this period, which spreads over redshifts $z \approx 5-15$, is crucial to learn more about the nature, timing and mechanism of the formation and evolution of the aforementioned astrophysical objects, as well as their impact on the interstellar (ISM) and intergalactic (IGM) medium surrounding them \citep{Ciardi_2005,Morales_2010,Pritchard_2012,Furlanetto_2016,Liu_2020}. From observations of the Gunn-Peterson troughs of high-$z$ quasars \citep{Becker_2001,Fan_2006} and the optical depth for Thomson scattering of the Cosmic Microwave Background (CMB) radiation \citep{Planck_2016}, we can deduce that most reionization took place in the range $6 \lesssim z \lesssim 10$, with recent observations suggesting an end of reionization at $z <6$ (see e.g. \citealt{Becker_2015} and \citealt{Bosman_2022}).

There exist multiple indirect probes to study this period. For example, the evolution of the observed Lyman-$\alpha$ emitter luminosity function at $z>6$ \citep{Clement_2012, Schenker_2013}, and Lyman-$\alpha$ absorption profiles to distant quasars \citep{Mortlock_2016,Greig_2017,Davies_2018}. However, the most sensitive probe to study the EoR is through the fluctuations of the redshifted 21-cm line of neutral Hydrogen against the CMB \citep{Hogan_1979,Madau_1997,Shaver_1999,Tozzi_2000,Ciardi_2003,Zaroubi_2013}. A statistical detection of the strength of these 21-cm brightness temperature fluctuations can allow us to constrain our models of the early Universe and the formation of the first stars and galaxies. For this, a number of interferometric low-frequency radio telescopes have been designed, such as LOFAR\footnote{Low-Frequency Array, \url{http://www.lofar.org}}, HERA\footnote{Hydrogen Epoch of Reionization Array, \url{https://reionization.org/}}, MWA\footnote{Murchison Widefield Array, \url{http://www.mwatelescope.org}} and PAPER\footnote{Precision Array to Probe EoR, \url{http://eor.berkeley.edu}}, which over the years have been providing increasingly tighter upper limits. For example, \citet{HERA_2022} reported $\Delta^2 (k = 0.34~h \rm Mpc^{-1}) \leq 457~ \rm mK^2$ at $z$=7.9 and $\Delta^2 (k = 0.36~h \rm Mpc^{-1}) \leq 3496~\rm mK^2$ at $z$=10.4 from 94 nights of observation, and \citet[][hereafter M20]{Mertens2020} reported $\Delta^2 (k = 0.075~h \rm Mpc^{-1}) < 5329~\rm mK^2$ from 141 hours ($\approx$10 nights) of observation with LOFAR at $z$=9.1. 

One of the major challenges faced in the detection of the 21-cm signal is the fact that it is buried under foregrounds (synchrotron and free-free emissions from the Milky Way and other galaxies) that are several orders of magnitude stronger. To address this issue, \citet{HERA_2022} uses the ``foreground avoidance" technique \citep{Kerrigan_2018,Morales_2019} by focusing on regions in Fourier space which are mostly foreground free, while the LOFAR EoR Key Science Project (KSP) team uses foreground modelling and removal, which allows the maximization of scales explored, as well as boosts the sensitivity up to an order of magnitude \citep{Pober_2014}. 

The most stringent constraints obtained with LOFAR data were presented in \citetalias{Mertens2020}, where Gaussian Process Regression \citep[GPR, as described in][]{Mertens2018,Gehlot_2019,Hothi_2021} was used for hyperparameter optimization with different Matern class functions (Eq.~\ref{eq:Maternfunc}, see below) chosen as covariance kernels for modelling different components of the observed data and then recovering the fitted datacube. However, \citet{Kern_2021} pointed out some issues with this approach. Primarily, they found that \revised{given the choice of normalization and bias correction in the power-spectra estimation used in \citetalias{Mertens2020},} misestimation of the covariance kernel for the EoR signal could lead to significant signal loss. This can have severe ramifications on the astrophysical interpretations of the estimated 21-cm signal power spectrum. \revised{Further, they show that alternative choices for normalization and weighting schemes could reduce the dependence on the choice of covariance priors, thus reducing its impact on the estimation of the 21-cm signal. However, here we focus on improving the covariance prior, while in future works we plan to explore other normalization and bias correction schemes to further upgrade the overall analysis pipeline.}

\revised{To improve the covariance kernel, we refer to \citet{Mertens_2023}. They} propose a Machine Learning (ML) based approach to GPR, where the covariance kernel for the 21-cm signal is obtained by implementing a machine learning based algorithm that learns from simulations. The results obtained can then be compared against runs of the same simulation code, to constrain the physical parameters used in it. As the covariance kernel is trained over a range of physical parameters, this would significantly reduce chances of misestimation, and thus it can be reliably used to derive astrophysical parameters necessary for the same simulation code to generate similar power spectra. 

In this paper, we use {\tt GRIZZLY} \citep{Ghara_2015,Ghara_2018,Ghara2020} for generating the training, test and validation datasets. This code has been employed previously \citep[see][]{Ghara2020} to constrain astrophysical parameters based on the results obtained in \citetalias{Mertens2020}. As this combines $N$-body simulations with 1D radiative transfer, it is more physically precise than semi-numerical algorithms, while not being as computationally expensive as codes that use 3D radiative transfer \citep[while still performing reasonably well, as shown in][]{Ghara_2018}.
Here, we test the performance of this ML based kernel versus covariance kernels used with GPR in previous work. In Section~\ref{sec:methodology}, we discuss a range of simulations used to generate mock 21-cm datasets, as well as introduce the ML trained 21-cm kernel. We also provide a short introduction to GPR. We report the results and comparisons between kernel performances in Section~\ref{sec:results}. Finally, we discuss the role of the excess noise component found in \citetalias{Mertens2020} and the overall performance of the ML based kernel in Section~\ref{sec:discuss}. In a companion paper we will apply the new pipeline to 10 nights of LOFAR data at $z$=9.1, as was done in \citetalias{Mertens2020}.

\section{Methodology}\label{sec:methodology}

In this section, we introduce the pipeline used to implement GPR to recover the 21-cm signal from mock datasets, comparing the performance of an ML based kernel versus kernels used in \citetalias{Mertens2020}.

\subsection{Simulations of the 21-sm signal}\label{sec:sims}

The 21-cm differential brightness temperature relative to the CMB for any patch of the IGM  is given by \citep[see][]{Furlanetto_2006eq}:
\begin{multline}\label{eq:dtb} \delta T_{ \rm b} = 27 x_{\rm HI} (1 + \delta_{\rm B}) \left(1 - \frac{T_{\rm CMB}}{T_{S}} \right) \\
        \times\biggl[ \left( \frac{\Omega_{\rm B}h^2}{0.023} \right) \left( \frac{0.15}{\Omega_{\rm m} h^2} \frac{1+z}{10} \right)^{1/2} \biggr]~\rm mK
\end{multline} where $x_{\rm HI}$ is the fraction of neutral hydrogen, $\delta_{\rm B}$ is the fractional overdensity of baryons, $T_{\rm S}$ is the spin temperature, $T_{\rm CMB}$ is the temperature of the CMB photons at that redshift, $\Omega_{\rm m}$ is the total matter density, $\Omega_{\rm B}$ is the baryon density, $z$ is the redshift and $h$ is the Hubble constant in units of 100 kms$^{-1}$Mpc$^{-1}$. In this equation, the parameters affecting the large-scale fluctuations of $\delta T_{\rm b}$ are $x_{\rm HI}$, $T_{\rm S}$ and $\delta_{\rm B}$.

We consider a variety of simulations to generate mock 21-cm differential brightness temperature maps as discussed below.  In Section~\ref{sec:GRIZZLY}, we employ maps generated using {\tt GRIZZLY}, where we focus on variations tied to fluctuations in $x_{\rm HI}$ and $T_{\rm S}$, while assuming that the fluctuations due to $\delta_{\rm B}$ are small and can thus be ignored. In Section~\ref{sec:MBII_Crash}, we do not make this assumption and employ maps generated using the reionisation simulation code {\tt CRASH}. In Appendix~\ref{appendix} we also consider the additional case of using {\tt 21cmFAST} \citep{Mesinger2007, Greig2015} to generate the 21-cm differential brightness temperature maps.

\subsubsection{{\tt GRIZZLY} simulations}\label{sec:GRIZZLY}

{\tt GRIZZLY} \citep{Ghara_2015,Ghara_2018,Ghara2020} employs a 1D radiative transfer scheme in combination with cosmological density fields and halo catalogues obtained from an N-body simulation to produce brightness temperature maps of the 21-cm signal at different redshifts for a given source model. The algorithm has been shown to reproduce results similar to those obtained with 3D radiative transfer schemes with the same N-body simulation, while being at least $10^5$ times faster \citep{Ghara_2018}. Because of this, we can run a large number of  {\tt GRIZZLY} simulations without the process being too computationally expensive. Furthermore, it has a wide range of physical parameters that can be varied, thus allowing us to explore the role of different physical processes in generating different models of the 21-cm signal. The density fields, velocity fields and the halo lists used in this work are obtained from the same N-body simulation (500 $h^{-1}$ cMpc box length, 6912$^3$ particles, with a mass resolution of $4.05 \times 10^7$~ \rm M$_{\rm \odot}$) which was used in \citet{Ghara2020}. In this study, we consider two major {\tt GRIZZLY} models presented in Sections 3.1 and 3.2 of \citet{Ghara2020}. Similar to their implementation, we use four physical parameters to model the sources: the ionization efficiency $(\zeta)$, the minimum mass of the UV emitting halos $(M_{\rm min})$, the minimum mass of the X-ray emitting halos $(M_{\rm min_{\rm X}})$ and the X-ray heating efficiency $(f_{\rm X})$. The emission rate of ionizing photons and X-rays per unit stellar mass from a halo are $\zeta\times 2.85\times 10^{45} ~ \rm s^{-1} M^{-1}_{\rm \odot}$ and $f_{\rm X}\times 10^{42}~ \rm s^{-1} M^{-1}_{\rm \odot}$, respectively. Further, the X-ray spectral index $\alpha$ is fixed at 1.2, as done in \citet{Ghara2020}. Lastly, they note that all other IGM properties can be derived from these parameters, and thus using just these to define the simulation is sufficient. The properties of the two models adopted are listed below:
 
\begin{itemize}
    \item {\bf $x_{\rm HI}$ fluctuation dominated model:} here, we assume a uniform Ly$\alpha$ background strong enough to allow the spin temperature $T_{\rm S}$ to be fully coupled to the gas temperature $T_{\rm K}$. Further, we adopt the following parameters: $\zeta = 7.0$, $M_{\rm min} = M_{\rm min_{\rm X}} = 10^9 \rm M_{\rm \odot}$ and f$_{\rm X} = 100$, which makes the gas temperature (and in turn, $T_{\rm S}$) significantly high compared to  $T_{\rm CMB}$ due to strong X-ray heating. This assumption of $T_{\rm S} \gg T_{\rm CMB}$ ensures that $\delta T_{\rm b}$ becomes insensitive to the $ (1 - \frac{T_{\rm CMB}}{T_{\rm S}})$ term from Equation~\ref{eq:dtb}. Thus, all variability of $\delta T_{\rm b}$ is tied to the fluctuation of the neutral hydrogen fraction $x_{\rm HI}$. 
    
    \item {\bf $T_{\rm S}$ fluctuation dominated model:} in this case, while we continue to have the assumption of a strong, uniform Ly$\alpha$ background to ensure coupling of $T_{\rm S}$ and $T_{\rm K}$, we change our parameters to relax the condition of $T_{\rm S} \gg T_{\rm CMB}$. This is done by reducing the X-ray heating and ionization efficiency. Thus, we adopt the following parameters: $\zeta = 3.0$, $M_{\rm min} = 10^9 \rm M_{\rm \odot}$, $M_{\rm  min_{\rm X}} = 10^{10} \rm M_{\rm \odot}$ and $f_{\rm X} = 1$. This allows for greater variability tied to $T_{\rm S}$, with regions of partial reionization and heating forming in the IGM.
\end{itemize}

\subsubsection{ {\tt CRASH} simulations}\label{sec:MBII_Crash}

As a reference, we also use the simulations of reionization described in \citet{Eide_2018,Eide_2020} and \citet{Ma_2021}. These are obtained by post-processing the large-scale, high-resolution hydrodynamical simulation Massive Black-II (\citealt{Khandai_2015}; box length $100~h^{-1}$cMpc, $2 \times 1792^3$ gas and dark matter particles, corresponding to a resolution of $2.2 \times 10^6 h^{-1} \rm M_{\rm \odot}$ and $1.1 \times 10^7 h^{-1} \rm M_{\rm \odot}$ respectively) with the multi-frequency 3D radiative transfer code {\tt CRASH} \citep{Ciardi_2001,Maselli_2009,Graziani_2013,Graziani_2018,Glatzle_2019}. Here, we make use of the ``Stars'' simulation run (which includes only stellar type sources) to generate the mock 21-cm signal data at $z$=9.18.
We refer the reader to the original papers for more detailed information on the simulations.

\subsection{Gaussian process regression}\label{sec:GPR}

Gaussian Process Regression (GPR, \citealt{RasmussenBook,Aigrain_2023}) can be used to model a noisy observation $\mathbf{y} = f(\mathbf{x}) + \epsilon$, with $\epsilon$ Gaussian noise having variance $\sigma_{\rm noise}^2$.  This is achieved by modelling the Gaussian Process as a joint probability distribution for {\bf y} = $\{y_i\}_{i=1,...,N}$, as \textbf{f}=$f$(\textbf{x}), which is fully defined by its mean vector (\textit{m}) and covariance matrix (\textit{K}, also called covariance ``kernel'') as: \begin{equation}\label{eq:gp}
    \mathbf{f} \sim \mathcal{N} ( m(\mathbf{x}), K (\mathbf{x},\mathbf{x}) ) \, ,
\end{equation} for a set of points {\bf x} (independent parameters). Here, the covariance matrix \textit{K} gives the covariance between the function values at any two points and can be written as $K_{ij} = \kappa(x_i,x_j,\phi) + \delta_{ij}\sigma_i^2$, where $\kappa(x_i,x_j,\phi)$ can be optimised by the choice of hyperparameters represented by $\phi$, and $\delta_{ij}$ is the Kronecker-delta function.

When applying it to radio data to extract the 21-cm signal, we split this function into a foreground component, $f_{\rm fg}$, and the 21-cm signal, $f_{\rm 21}$, giving: \begin{equation}\label{eq:basiceq}
y = f_{\rm fg}(x) + f_{\rm 21}(x) + \epsilon
\end{equation} where $y$ is the observed data and $x$ is the frequency. Next, following \citetalias{Mertens2020}, we further split the foreground component into the intrinsic sky emission component ($f_{\rm sky}$), which comes from the confusion-limited extragalactic sources and from the Milky Way, and the mode-mixing contaminants component $f_{\rm mix}$, which has contributions from the instrument chromaticity and calibration errors. 
Beyond the foreground, we also model the noise (represented by $\epsilon$ in Equation~\ref{eq:basiceq}) using estimates of the noise variance for $\approx$10 nights of observation from \citetalias{Mertens2020}. In addition to this, \citetalias{Mertens2020} found a significant spectrally-correlated residual, and thus we inject this ``excess noise'' component ($f_{\rm ex}$) into our model as well. This gives an updated version of Equation~\ref{eq:basiceq}: \begin{equation}\label{eq:fulleq}
    y = f_{\rm sky}(x) + f_{\rm mix}(x) + f_{\rm ex}(x) + f_{\rm 21}(x) + \epsilon \, .
\end{equation}

For the sake of simplicity and utilising the additive property of matrices, we can reduce Equation~\ref{eq:fulleq} to $y = f(x) + \epsilon$, and represent $f(x)$ with its corresponding covariance kernel $K$ (from Equation~\ref{eq:gp}) as: 
\begin{equation}\label{eq:kerneq}
    K = K_{\rm sky} + K_{\rm mix} + K_{\rm noise} + K_{\rm ex} + K_{\rm 21} \, .
\end{equation}

\citetalias{Mertens2020} modelled each of these kernels using the best possible fit Matern-class functions \citep[Eq.~\ref{eq:Maternfunc};][]{Stein1999}:

\begin{equation}\label{eq:Maternfunc}
    K_{\rm Matern}(r) = \sigma^2 \frac{2^{1-\eta}}{\Gamma (\eta)} \left( \frac{\sqrt{2\eta }r}{l} \right)^{\eta} \kappa_{\eta} \left( \frac{\sqrt{2\eta}r}{l} \right).
\end{equation}
Note that in the Matern-class function, $r$ is the absolute difference between the frequencies of two sub-bands, $\kappa_{\eta}$ is the modified Bessel function of the second kind and $\Gamma$ is the Gamma-function. 

\citetalias{Mertens2020} obtained the best possible fit Matern-class function by taking different values of the hyperparameter $\eta$, maximising the marginal likelihood (also known as the evidence) and obtaining estimates for the coherence scale hyperparameter $l$ and the variance $\sigma^2$. Then, for each kernel, they chose the $\eta$ that led to the highest evidence by calculating the analytical integral over \textbf{f} which is the log-marginal-likelihood \citep[LML, see Section 2.3 in][]{Mertens2018} and choosing the kernel that maximises its value. For calculating the hyperparameters (listed in the second column of Table~\ref{table:results}), \citetalias{Mertens2020} used a gradient-descent based optimization algorithm for maximising the LML. 

\subsection{Machine learning trained 21-cm kernel}\label{sec:ML-GPR}

The limitations of GPR as pointed out by \citet{Kern_2021} mainly boil down to the choice of covariance kernel for the 21-cm signal. While the choice of hyperparameters allows a variety of functions to be accessed,  the same function might not work equally well across the $k$-space. Having a function obtained by employing ML trained on power spectra of simulations where the sources of ionization are modelled using parameters that sample a wide range of values, allows greater flexibility, and reduces chances of misestimation. Further, it allows for a direct comparison with physical quantities, as we can reliably derive the source parameters necessary to generate simulations that produce the power spectra estimated by the ML-based kernel.

To achieve this, \citet{Mertens_2023} uses a Variational Auto-Encoder \citep[VAE,][]{Kingma_2013,Kingma_2019} algorithm. Simply put, an Auto-Encoder \citep[AE,][]{Goodfellow_2016} is an unsupervised neural network which compresses data by reducing the number of independent parameters used to describe it into what is referred to as a ``latent space'' of hyperparameters. Thus, it is primarily used for data compression by filtering out independent parameters that are deemed to be unnecessary because they only slightly affect data recovery. This is a two steps process, where the first step of reducing the number of independent parameters into the latent space is called {\sl encoding}, while the step of recovering the data given the latent space parameters is called {\sl decoding}. Instead of taking an input of just a set of parameters $a_1,~...~,a_n$, a VAE \citep{Cinelli_2021} uses probability distributions of each parameter, thus allowing to interpolate in the latent space, and to generate a large range of new samples of reconstructed data (in our case 21-cm signal models), which are not limited by the data that the encoder was trained on. However, this also means that the VAE encoder has an inherently larger error than an AE encoder. 

Thus, reconstructing the training data using the decoder with the latent space generated by the VAE encoder as input would not be an exact match to the original training data. However, the VAE is designed to optimize a trade-off between reconstruction accuracy and the fidelity of the latent space representation, by minimising the KL (Kullback-Leibler) divergence loss \citep{Kullback_1951}, which is a measure of the divergence between the distribution of reconstructed data and the training data. So while the reconstructed data and training data might not be an exact match, if their overall distribution is similar (i.e., divergence is minimised), the training is considered successful. Thus, in the training of a VAE algorithm, the following sources of error exist: \begin{enumerate}
    \item {\sl Encoder:} the error due to sampling from a distribution for each independent parameter to build a latent space. Sampling from a distribution is expected to be noisier than choosing point values.
    \item {\sl Decoder:} the error due to deriving the independent parameters, given some value of the latent space parameters. As this does not involve sampling from distributions, the contribution to the total error is expected to be smaller.
\end{enumerate} Mertens et al. (2023) shows that GPR can be used to estimate the values of the latent space parameters, after which the decoder of the VAE kernel can be used to estimate the independent parameters, and in turn to obtain the recovered 21-cm signal. In this case we would then use only the decoder part of the VAE algorithm. However, as minimising KL divergence loss requires to train the encoder and decoder together, we proceed as follows.

We start by using two hyperparameters, $x_1$ and $x_2$, and an associated variance, as the latent space parameters. We train the VAE algorithm on a dataset of $\approx1500$ simulations (the training set), and validate it against an independent dataset of $\approx150$ simulations (the validation set). We refer to this as the {\sl VAE kernel}. The training and validation datasets are generated by running {\tt GRIZZLY} simulations with a wide range of values for the parameters introduced in Section~\ref{sec:GRIZZLY}. We sample them randomly at the same redshift as the targeted 21-cm signal ($z$ = 9.16, 8.30 and 10.11) and in the following ranges: $\zeta = [-3, 3]$, $M_{\rm min} = [9, 12]$, $M_{\rm min_{\rm X}} = [9, 12]$ and $f_{\rm X} = [-3, 2]$ in the log space. \revised{We choose these ranges to be significantly broader than necessary (i.e., the performance of the VAE kernel does not show any appreciable difference even if they are reduced by multiple orders of magnitude), and also note that the performance of the kernel is not impacted if we fix one of the parameters to a specific value and vary the remaining three. This ensures that the sampling range does not induce any major bias. Next, to} train the VAE we use 2000 iterations, as we find that the KL divergence loss and the reconstruction loss for both the training and validation datasets stabilise after $\approx$500 iterations. 
The reconstruction loss is defined as the total error made when using the encoder to obtain specific values of the latent space parameters, and then employing the decoder to retrieve the data from those parameters. 
We then evaluate the ratio between the output and input data as a function of the wave number $k$, and finally calculate the median ratio, which is $\approx$1 across all $k$-bins. However, as discussed earlier, the 68\% confidence interval is significant, being $\approx$10\% for $k = [0.06, 0.12] ~h \rm Mpc^{-1}$, $\approx$35\% for $k= [0.12, 0.43] ~h \rm Mpc^{-1}$ and $\approx$27\% for $k= [0.43, 1.11] ~h \rm Mpc^{-1}$ at $z$= 8.30, 9.16 and 10.11.

We note a similar reconstruction error for training sets of sizes ranging from 1000 to 5000 simulations (with the validation set scaling as $\sim10\%$ of the training set in size), while the errors become worse when using less than 1000 simulations. We also tried to used three hyperparameters, but saw no significant improvement in the recovery error. We thus choose to use two hyperparameters to avoid overfitting. 

A high reconstruction error was to be expected, as it includes also the error due to the encoding process. While this is not required for our purpose, we do need to evaluate the exact contribution from the decoder. For this, we create a testing set of $\approx150$ simulations, and ensure that it also includes ``extreme'' models from GRIZZLY (which we define as cases where the power spectrum differs by at least an order of magnitude from the mean of the power spectra from the training and validation datasets), along with some injected stochastic noise. We then apply GPR for hyperparameter optimization (using MCMC, as discussed in Section~\ref{sec:recovery}) to estimate $x_1$, $x_2$ and the associated variance. This is then used with the decoder to obtain the recovered signal. The median ratio of the recovered and input data across all $k$-bins is again $\approx$1. But now we obtain a 68\% confidence interval of  $\approx$0.5\% for $k = [0.06, 0.12] ~h \rm Mpc^{-1}$, $\approx$1\% for $k = [0.12, 0.43] ~h \rm Mpc^{-1}$ and $\approx$3\% for $k = [0.43, 1.11] ~h \rm Mpc^{-1}$ at $z $= 8.28, 9.16 and 10.11, respectively. By comparing to the total error estimated above, we note that even where decoder error is high, i.e., for the highest $k$-bins, it is still a minor contributor. Therefore, as the overall error is $\lesssim 5\%$, we accept the VAE kernel as a reliable ML-based kernel for the 21-cm signal, and proceed to use it with GPR for signal recovery.

\subsection{Generating mock datasets}\label{sec:genmock}

To build a mock dataset in the gridded $U-V$ cube domain of radio observations, we derive a full dataset $y$ by adding each term on the right hand side of Equation~\ref{eq:fulleq}. For this, we adopt the values of $\eta$, $l$ and $\sigma^2$ for $K_{\rm sky}$, $K_{\rm mix}$ and $K_{\rm ex}$ given in the second column of Table~\ref{table:results} to generate their corresponding $f_{\rm sky}$, $f_{\rm mix}$ and $f_{\rm ex}$. These values are obtained from the results of \citetalias{Mertens2020}, where they used $\eta_{\rm sky} = +\infty$ for the intrinsic sky, $\eta_{\rm mix} = 3/2$ for the mode-mixing contaminants, and $\eta_{\rm ex} = 5/2$ for the excess noise to fix Matern class functions for each of these components, and then employed GPR to obtain the coherence-scale hyperparameter $l$ and its associated variance by adopting $\eta_{\rm 21} = 1/2$ for the 21-cm signal. \revised{Note that while $f_{\rm sky}$ and $f_{\rm mix}$ are not independent quantities, their recovery with GPR treats them as such. In this paper, we also generate them as independent components to build our mock datasets (unlike in real data, where the mode-mixing component depends on the true sky signal). Thus, the overall quality of the recovered $f_{\rm sky}$ and $f_{\rm mix}$ is better from the mock datasets, than from real data. However, the effect on the accuracy of recovery is not expected to be severe, as the impact of not factoring in the inter-dependency is insignificant, as compared to the overall power of these components. That is, even without assuming their inter-dependence, GPR can recover them with reasonably high accuracy.}

Next, we also assume the value $\sigma_{\rm noise}^2 \sim 74 \times~10^3$ mK$^2$ \citepalias[from][]{Mertens2020} to simulate the noise component $\epsilon$ for $\approx$10 nights of observation with LOFAR scaled to the Field of View (FoV) corresponding to the {\tt GRIZZLY} simulations, which is $3.03^{\circ} \times 3.03^{\circ}$. The variance for the intrinsic sky-emission, mode-mixing and excess noise components scales by the noise variance as indicated in Table~\ref{table:results}.

We also consider the case of $\approx$100 nights of observation, to provide estimates for future observations of similar duration with LOFAR. In this case, the noise variance is expected to be reduced by a factor of 10, assuming all effects to scale uniformly from $\approx$10 to $\approx$100 nights of data. However, it should be noted that the data used in \citetalias{Mertens2020} was plagued with issues such as RFI flagging and bad ionospheric conditions, and it would be of little scientific value to assume the same conditions to persist for a 10 times longer observational duration. For this reason, we assume ideal conditions (for example, picking observation nights with good ionospheric conditions) and assume that the thermal noise is reduced by a factor of 20 instead, obtaining $\sigma_{\rm noise}^2 \sim 3.7~\times~10^3~\rm mK^2$. 
\revised{As we expect the excess noise to behave in a similar fashion, we also reduce its variance $\sigma_{\rm ex}^2$ by a factor of 20. \citetalias{Mertens2020} noted that the mode-mixing contaminants were not decreasing when integrating over more nights of data. However, the parts of this component due to effects of ionosphere and calibration errors are uncorrelated from night to night, and thus are expected to decrease with longer integration, leading to a reduction of the mode-mixing term. Additionally, this should also decrease because of the improved UV-coverage. Here, we thus consider an ideal scenario to understand how the performance of the VAE kernel compares with previously used kernels when also the mode-mixing term's variance $\sigma_{\rm mix}^2$ is reduced by a factor of 20.} However, we scale down the intrinsic sky component only by a factor of 2. This is justified because ({\it i}) the confusion noise limit is set by the resolution of the LOFAR telescope, so that unresolved point sources cannot be subtracted even when integrating over more nights; 
({\it ii}) modelling an increasing number of point sources is not a trivial task, and doing it accurately does not seem feasible in the short-term. Thus, we limit ourselves to an assumption of modest improvements. Because of this, we assume that the variance of $f_{\rm sky}$ is scaled by a thermal noise variance of $\sigma_{\rm noise}^2 \sim 37~\times~10^3~\rm mK^2$, corresponding to a factor of 2 reduction from the noise variance in the $\approx$10 nights case.

Following the procedure outlined above, we generate two sets of foreground, noise and excess noise components, i.e. for $\approx$10 and $\approx$100 nights of observation. We then add the 21-cm signal (see Section~\ref{sec:sims}) to obtain a complete mock dataset.

\begin{figure}
\centering
\includegraphics[width=0.95\columnwidth,keepaspectratio]{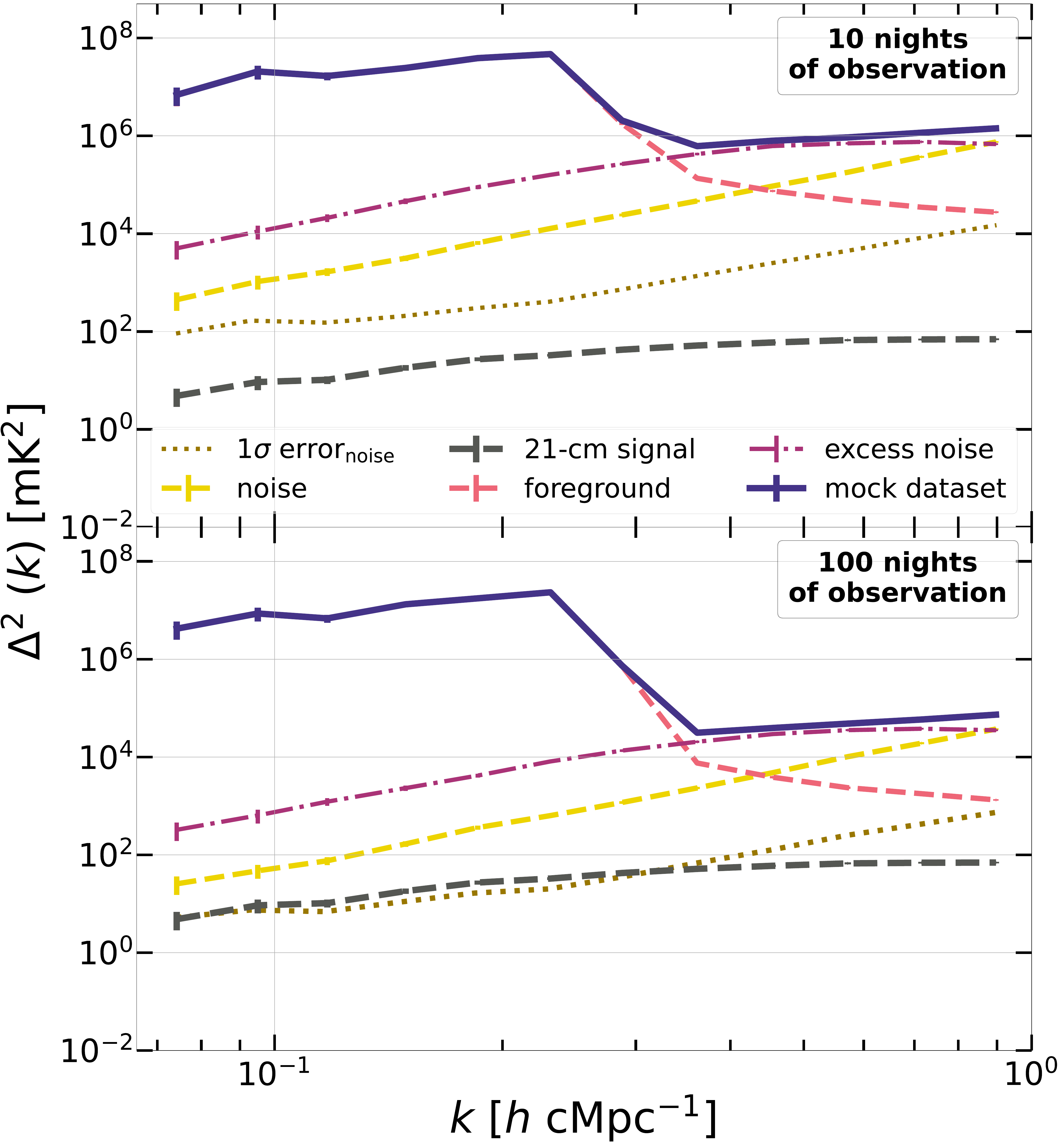}
\caption{Power spectrum of the mock dataset (purple solid line) generated using the $x_{\rm HI}$ fluctuation dominated model from {\tt GRIZZLY}. It consists of the foreground component (intrinsic sky + mode-mixing contaminants; pink dashed), the excess noise (magenta dashed), the noise (yellow dashed) and the 21-cm signal at $z = 9.16$ (grey dashed). We also show the 1 - $\sigma$ upper limit (dark-yellow dotted) achievable if the dataset is thermal noise dominated, i.e. any signal below this line has SNR < 1. The top (bottom) panel refers to a case with the noise corresponding to 10 (100) nights of observation.} 
\label{fig_inputUTs}
\end{figure}

\begin{figure}
\centering
\includegraphics[width=0.95\columnwidth,keepaspectratio]{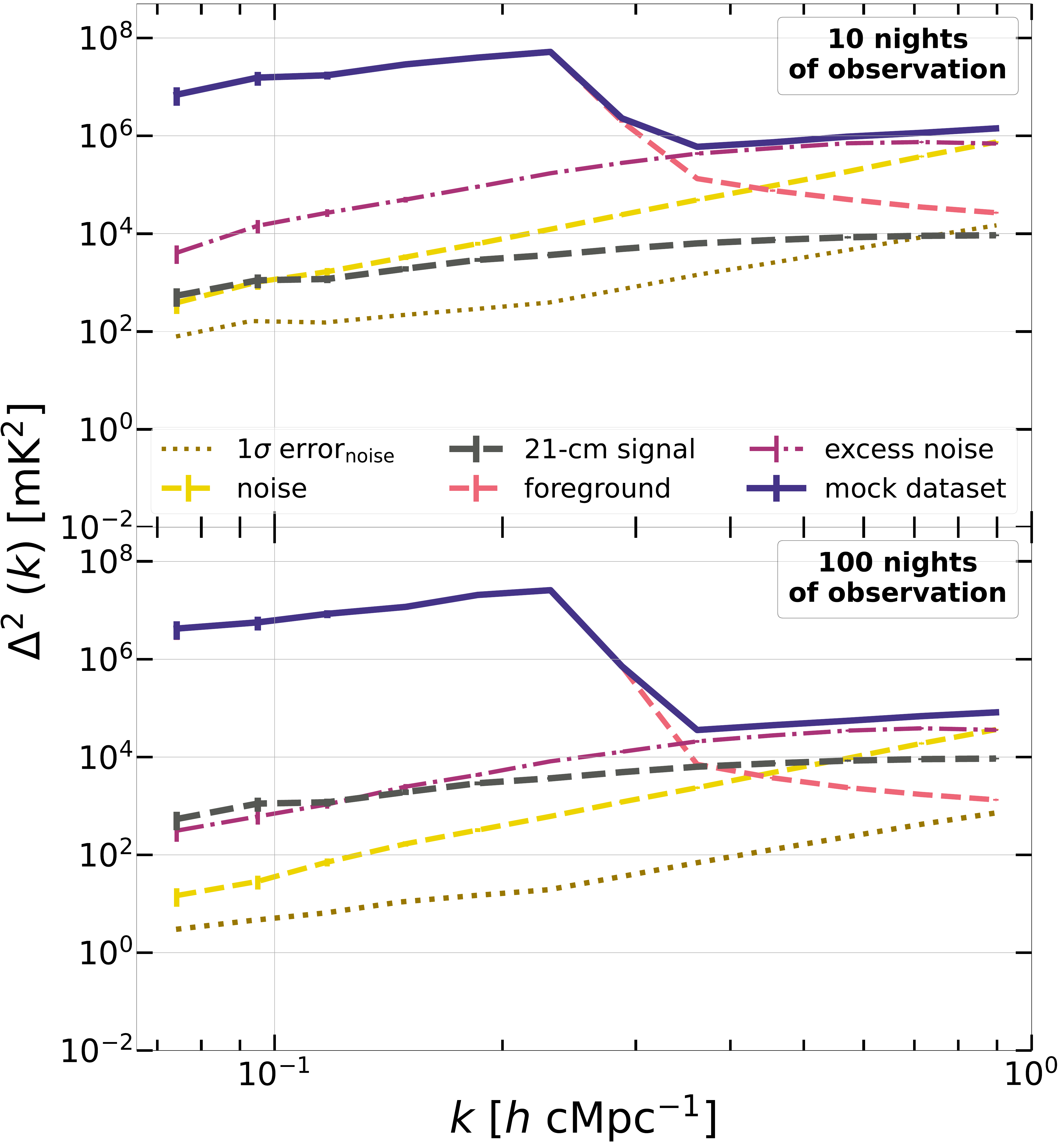}
\caption{As Fig.~\ref{fig_inputUTs} for the $T_{\rm S}$ fluctuation dominated model. }  
\label{fig_inputVTs}
\end{figure}

The input power spectra for the overall mock dataset using {\tt GRIZZLY}  is shown in Figures~\ref{fig_inputUTs} and~\ref{fig_inputVTs} for the $x_{\rm HI}$ and $T_{\rm S}$ fluctuation dominated case, respectively. The ratio between the 21-cm signal and the 1-$\sigma$ uncertainty of the noise power spectrum is a measure of the Signal to Noise Ratio (SNR). Thus, for $k$-bins where the 21-cm signal's power is greater than the 1-$\sigma$ error of the noise, there is a chance of detectability with 1-$\sigma$ confidence. From Figure~\ref{fig_inputUTs} we can thus conclude that for $\approx$10 nights of observation a detection of the $x_{\rm HI}$ fluctuation dominated 21-cm signal is unlikely, as the SNR is $\ll1$ for all $k$-bins. However, the chances of detectability improve for $\approx$100 nights of observation, as SNR $\approx1$. 
In Figure~\ref{fig_inputVTs}, we see that detecting the $T_{\rm S}$ fluctuation dominated signal should be possible also for $\approx$10 nights of observations, as SNR is $>1$ for all $k$-bins, while for $\approx$100 nights it is $\gtrsim10$, assuring a detection as long as the covariance kernel used to model the 21-cm signal is correctly estimated.

While we first focus on the 21-cm signal at $z = 9.16$ in order to make a direct comparisons with \citetalias{Mertens2020}, we also consider $z = 8.30~ \text{and}~10.11$ to prepare for future LOFAR results at these redshifts.

\subsection{Recovery with MCMC}\label{sec:recovery}

As discussed in Section~\ref{sec:GPR}, \citetalias{Mertens2020} used a gradient-descent method to maximise the LML. However, in this paper we instead use MCMC sampling \citep{Foreman_Mackey_2013} to estimate the hyperparameters by sampling their posterior distributions. The benefit of MCMC sampling is that it allows us to also have a measurement of the uncertainty on the hyperparameters, which can then be propagated. We build the posterior distributions adopting flat uniform priors with broad ranges as used in \citetalias{Mertens2020}. For the coherence-scales, we provide a smaller range for the uniform prior ($U$) to improve convergence time, as done by \citetalias{Mertens2020}. Thus, as listed in the third column of Table~\ref{table:results}, we use a range of (10,100) for $l_{\rm sky}$ and of (1,10) for $l_{\rm mix}$. 
To ensure that the converged value  for $l_{\rm ex}$ remains in the 1-$\sigma$ confidence interval, we used a range of (0.1, 0.8) rather than of (0.2,0.8) as in \citetalias{Mertens2020}. While we note that a Gamma prior for the variances of the different component leads to faster convergence, we still adopt flat priors in the logarithmic scale over several orders of magnitude (thus effectively an uninformed prior) to minimise the chances of bias.

In addition to this, we note that the coherence scale and the variance for the 21-cm signal are dependent on the baseline length. 
So, while to recover the intrinsic sky, mode mixing and excess noise components we continue to employ the same Matern class functions used to generate them (i.e. with the same $\eta$ values of the input, see Table~\ref{table:results}),
when using Matern class functions to recover the 21-cm signal, we modify the hyperparameter $l$ and the variance as: \begin{equation}
    l = \frac{l_0}{1+0.001l_{\alpha}l_0 (u - u_{\rm min})} \\\, \rm and \\\,
    \sigma^2 = \sigma^2_0 \sigma^2_{\rm norm} \left(\frac{u}{u_{\rm min}}\right)^{\sigma^2_{\alpha}},
\end{equation} 
where $l_0$ and $\sigma^2_0$ are the coherence-scale parameter and variance used in \citetalias{Mertens2020} for baseline length $u$ (where $u_{\rm min}$ is the minimum baseline length), but now we introduce the additional parameters $l_{\alpha}$ and $\sigma^2_{\alpha}$ to fully define the coherence-scale hyperparameter and the variance. Further, $\sigma^2_{\rm norm}$ is chosen such that the mean of the variance over all baselines is $\sigma^2_0$. The two additional parameters (i.e., $l_{\alpha}$ and $\sigma^2_{\alpha}$) thus allow us to encode the dependence on the baseline length into the covariance kernel for the 21-cm signal.

To recover the 21-cm signal component, we use Matern class functions with two specific values of $\eta$: the Exponential Matern class function with $\eta = 1/2$ (which, as shown in \citetalias{Mertens2020}, maximises the evidence), and the Matern32 function with $\eta = 3/2$. We then  compare their performance for recovery to the VAE kernel using GPR. For the hyperparameters $x_1$ and $x_2$, we again take an uninformed flat prior \revised{in the linear space}.

\section{Results}\label{sec:results}

Here we discuss a qualitative comparison between the results which are shown in Figures~\ref{fig_outputUTs} and~\ref{fig_outputVTs} using the three aforementioned kernels (Exponential, Matern32 and VAE). We then analyse the estimated coherence-scale hyperparameter and variance values for each of the components of the mock datasets defined in Section~\ref{sec:GRIZZLY} in Table~\ref{table:results}. Further, we qualitatively compare the results obtained by using the full simulations of reionization (Section~\ref{sec:MBII_Crash}) rather than the mock datasets generated with {\tt GRIZZLY}. Lastly, we explore the role of redshift on the performance of the VAE kernel, by testing cases at $z = 8.30$ and 10.11, and by comparing them against the results obtained for $z = 9.16$.

\subsection{$x_{\rm HI}$ fluctuation dominated model} \label{sec:uniformTS_results}

\begin{figure}
\centering
\includegraphics[width=0.95\columnwidth,keepaspectratio]{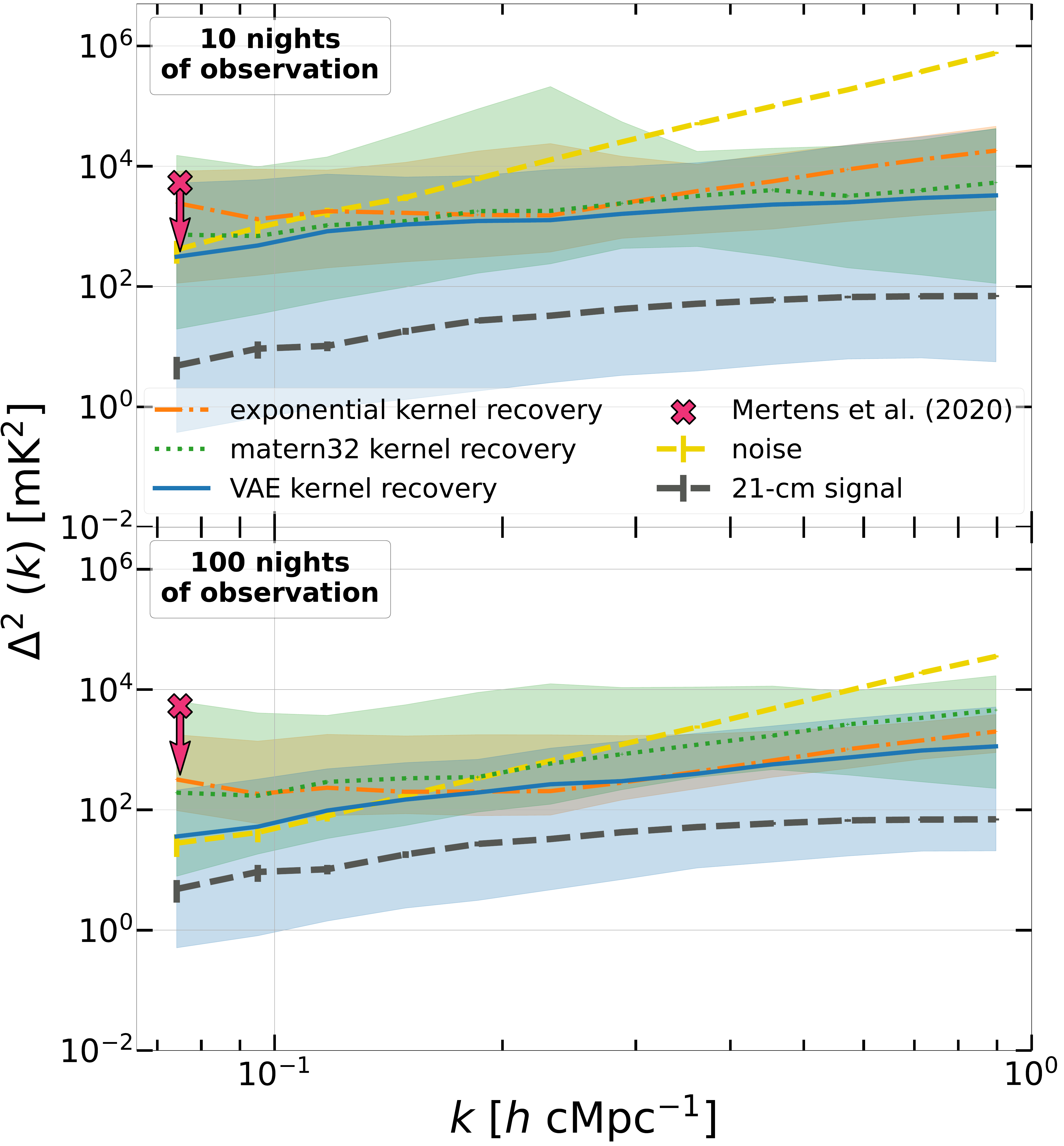}
\caption{Power spectrum recovered using the Exponential (orange dashed-dotted line) and Matern32 (green dotted) Matern class functions based covariance kernels, and the VAE-based kernel (blue solid), together with the $x_{\rm HI}$ fluctuation dominated model 21-cm signal (grey dashed) and noise (yellow dashed) at $z = 9.16$. The 2$\sigma$ uncertainty on the recovered signal for each kernel is shown as a shaded area in the corresponding colours. The top (bottom) panel refers to a case with the noise corresponding to 10 (100) nights of observation.
We also plot the estimated upper limits of power at $k = 0.075 ~h \rm Mpc^{-1}$ from \citet[][cross]{Mertens2020}. Note that this value can be significantly higher than the signal due to more complex noise in real data.
}
\label{fig_outputUTs}
\end{figure}

In Figure~\ref{fig_outputUTs}, we compare the results from the three kernels \revised{(VAE in blue solid, Exponential in orange dashed-dotted and Matern32 in green dotted) in recovering the power spectrum of the 21-cm signal at $z = 9.16$ (in grey dashed). We also show the 2-$\sigma$ uncertainty on the recovery from different kernels to compare their performance. We note that if the lower bound of the 2-$\sigma$ uncertainty of the recovered signal is below the uncertainty on the thermal noise (as shown in Figure~\ref{fig_inputUTs}) the recovery qualifies as an upper limit, otherwise it is referred to as a detection. Based on this, we note that in this case, the recovery from all kernels is going to provide upper limits, as the thermal noise uncertainty is higher than the 21-cm signal for $\approx$10 nights, and comparable to it for $\approx$100 nights.}

We note that for $\approx$10 nights of observations, while the VAE kernel has uncertainty bands wider than the Matern-class function based kernels,  the input 21-cm signal is contained  within its constrained region. Thus, while the recovered signal for the three kernels is comparable, the VAE kernel is robust enough to compensate for the overestimation \revised{and to contain the signal within the 2-$\sigma$ limits of the error, although it is still an upper limit rather than a constrained detection}. As discussed in Section~\ref{sec:genmock}, the reason for this non-detection when using Matern class functions based kernels and broad uncertainty bands for the VAE kernel is due to the low SNR, which is $<0.1$ across all $k$-bins. This, however, improves to an average SNR of $\approx1$ for $\approx$100 nights of observations, for which, as expected, we obtain tighter constraints and an improved prediction of the actual value. We see, though, that the recovered signal from Matern-class function based kernels is still over-predicted. In particular, for $k>0.5~h \rm cMpc^{-1}$ the Exponential and Matern32 kernels are unable to contain the signal even in their 2$\sigma$ uncertainty bands. On the other hand, the VAE kernel contains the signal in its 2$\sigma$ uncertainty bands across all $k$-bins, despite the recovered signal being about an order of magnitude higher than the input signal. The VAE kernel also does a much better job in recovering the overall shape of the power spectrum. By comparing to the estimated power at $k = 0.075 ~h \rm Mpc^{-1}$ from \citetalias{Mertens2020} (cross), it is clear that in this case the VAE kernel is also capable of significantly improving the 21-cm upper limits estimate. \revised{However, we highlight that the VAE kernel applied to real data is still likely to provide upper limits higher than what has been shown here, because of the more complex noise, and thus the improvement provided by the kernel might be lesser.}

In the 4$^{\rm th}$ and 5$^{\rm th}$ columns of Table~\ref{table:results} we show the MCMC estimates for the coherence-scale hyperparameter and the variance obtained by applying GPR to the input power spectrum of the mock dataset (in indigo, Figure~\ref{fig_inputUTs}). 
The covariance kernels for the intrinsic sky, mode mixing and excess noise components are the same as those used to generate them (i.e., the value of $\eta$ is fixed), while we adopt the VAE kernel for the 21-cm signal. Note that the variances for all components are scaled down by the corresponding value of $\sigma_{\rm noise}^2$. This is equal to $74~\times 10^3~\rm mK^2$ for all components for $\approx$10 nights of observations (see \citetalias{Mertens2020}), while for $\approx$100 nights this corresponds to scaling down by a factor of 2 (i.e., $\sigma_{\rm noise}^2 = 37~\times 10^{3}~\rm mK^2$) for the intrinsic sky component, and by a factor of 20 (i.e., $\sigma_{\rm noise}^2 = 3.7~\times 10^{3}~\rm mK^2$ ) for other components (as discussed in Section~\ref{sec:genmock}).
From the MCMC estimates, we note that the measurement of the coherence-scale for the $f_{\rm sky}$ and $f_{\rm mix}$ improves from $\approx$10  to $\approx$100 nights of observation.
However, the variance estimates do not show an improvement, and even slightly worsen for the excess noise component. The estimates of $x_1$ and $x_2$ and their associated variance $\sigma_{\rm 21}^2$ agree within error limits for both cases.

\subsection{$T_{\rm S}$ fluctuation dominated model} \label{sec:varTS_results}

\begin{figure}
\centering
\includegraphics[width=0.95\columnwidth,keepaspectratio]{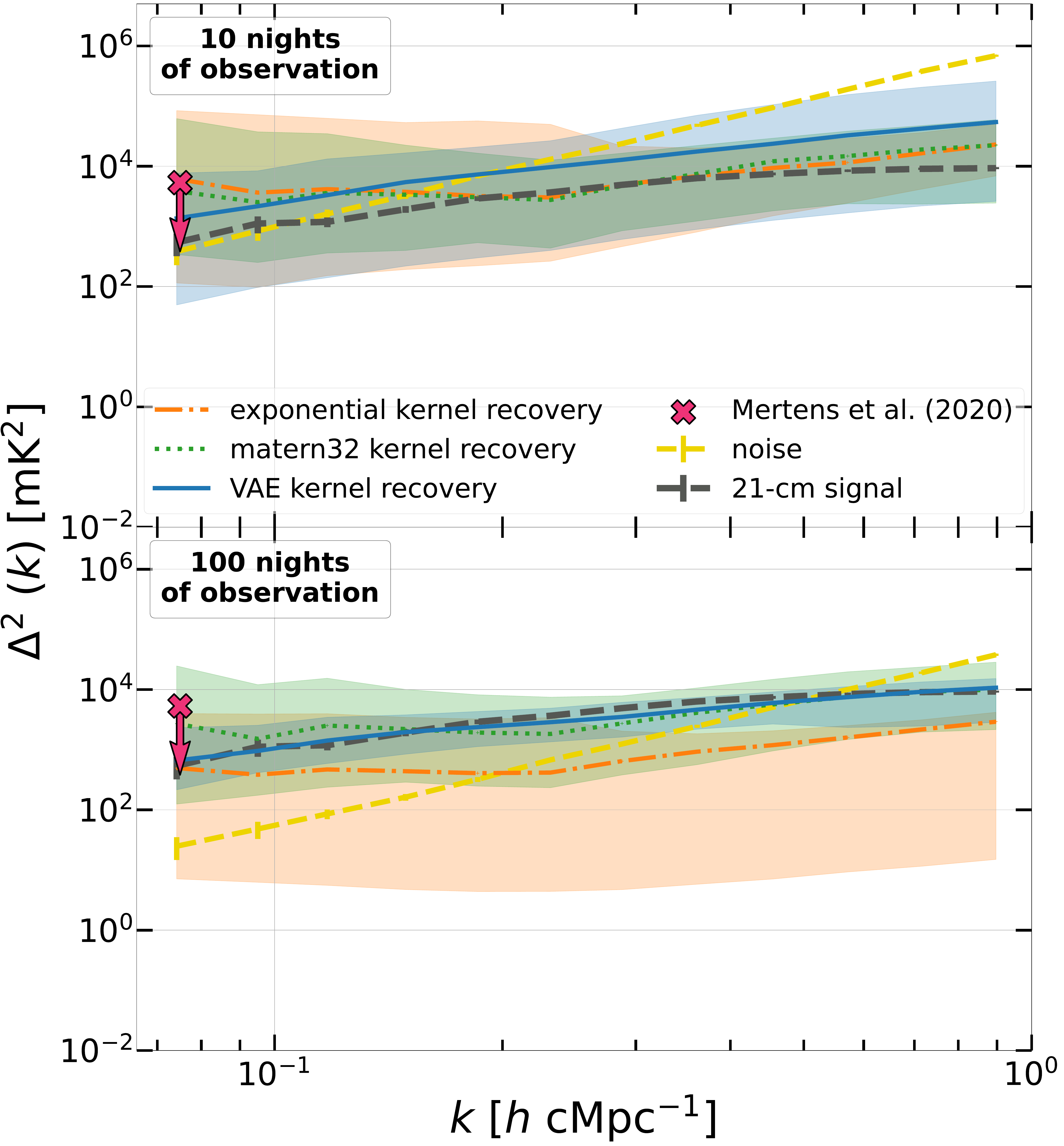}
\caption{As Figure~\ref{fig_outputUTs} for the $T_S$ fluctuation dominated model.}  
\label{fig_outputVTs}
\end{figure}

Here we perform a comparison between covariance kernels for the spin temperature fluctuation model, similarly to what done in the previous section. The results are shown in Figure~\ref{fig_outputVTs}. As seen in Figure~\ref{fig_inputVTs} and discussed in Section~\ref{sec:genmock}, the SNR is larger than $1$ also for  $\approx$10 nights of observations, suggesting better chances of detectability and smaller uncertainty ranges. Indeed, all three covariance kernels contain the signal within the 2$\sigma$ uncertainty bands around their recovered signal, and the uncertainty range for the VAE kernel is $\approx2$ orders of magnitude smaller than the equivalent case for the $x_{\rm HI}$ fluctuation dominated model. \revised{As discussed in the previous section, we compare the lower bounds of the recovered signal with the thermal noise uncertainty from Figure~\ref{fig_inputVTs} to classify the recovery as a detection or an upper limit.}

For $\approx$100 nights of observations, the SNR is $\gg10$ across all $k$-bins, so that the recovered signal is expected to reproduce the input signal with a significantly narrower 2$\sigma$ uncertainty range, provided that the covariance kernel chosen is a reliable estimate of the covariance of the input 21-cm signal. 
Indeed, from Figure~\ref{fig_outputVTs} we note that the VAE kernel fully recovers the signal with less than one order of magnitude uncertainty. However, the Exponential kernel \revised{contains} the 21-cm signal only in the lowest $k$-bins and shows significant bias in the estimated power at higher $k$-bins. \revised{Its broad 2-$\sigma$ uncertainty shows that the recovery just provides upper limits even in the low $k$-bins. On the other hand, the Matern32 kernel performs better, and provides a successful detection, albeit with broader uncertainty ranges than the VAE kernel recovery. This suggests that the Exponential kernel is definitely not a good match for the covariance of the input 21-cm signal, and, as highlighted by \citet{Kern_2021}, would lead to significant errors in the estimated physics, if used. The Matern32 kernel is better, however the VAE kernel improves upon it even further.} This problem \revised{with the Exponential kernel} appears in the $\approx$100 and not in the $\approx$10 nights of observation due to the similarity of power and shape of the excess-noise and 21-cm signal components. Thus, a covariance kernel which is not a reliable estimate of the covariance of the input 21-cm signal would not be able to distinguish between the two, and may either ignore both equally, or identify one over the other purely by chance. It can also be argued that the only reason for any ``detection'' at low $k$-bins using the Exponential kernel could possibly just be the detection of the excess noise component, wrongly interpreted as the 21-cm signal one.

The recovered values for the coherence-scales and variances for the intrinsic sky, mode mixing and excess noise components, as well as those for the hyperparameters $x_1$ and $x_2$ and associated variance for the 21-cm signal are listed in the 6$^{th}$ and 7$^{th}$ columns of Table~\ref{table:results} along with their 68\% confidence intervals. As expected, we find an improvement in recovery of the input values for $\approx$100 nights of observation in comparison to $\approx$10 nights, particularly for the coherence-scale hyperparameter. 

While we note better estimates for $l_{\rm sky}$ and $\sigma_{\rm sky}^2/\sigma_{\rm noise}^2$ for $\approx$100 nights of observations in both 21-cm signal models, in the $x_{\rm HI}$ fluctuation dominated model the input $\sigma_{\rm sky}^2/\sigma_{\rm noise}^2$ is not included within the estimate error of the recovered values. A similar behaviour is observed in the recovery of the variance for $f_{\rm mix}$ and $f_{\rm ex}$ in the $x_{\rm HI}$ fluctuation dominated model for $\approx$100 nights of observations. Lastly, we also note that the $x_1$ and $x_2$ hyperparameters and associated variance for the 21-cm signal in both models agree within the error estimates. While the estimated variance for  $\approx$100 nights of observations is higher, it also has a broad 68\% confidence interval.

\begin{table*}
\centering
\caption{We use the results of \citetalias{Mertens2020} as our input (2$^{nd}$ column) for the hyperparameters (1$^{st}$ column) to generate a mock dataset composed of the $f_{\rm sky}$, $f_{\rm mix}$ and $f_{\rm ex}$ components, in addition to the $x_{\rm HI}$ ($T_{\rm S}$) fluctuation dominated models from {\tt GRIZZLY} to generate the 21-cm signal component at $z = 9.16$. 
For the recovery using GPR, we keep the value of $\eta$ fixed for the $f_{\rm sky}$, $f_{\rm mix}$ and $f_{\rm ex}$ components, and estimate the median and the 68\% confidence intervals for the coherence-scales and variances from the MCMC based LML maximization (4$^{th}$ to 7$^{th}$ columns) using the priors as in \citetalias{Mertens2020} (3$^{rd}$ column), i.e. we either use linear scale uniform priors $U$, or logarithmic scale indicated as --. As the latter is over several orders of magnitude, it has not been listed and can be assumed to be an uninformed prior from $-\infty$ to $+ \infty$.
For $\approx$10 nights of observation we use $\sigma_{\rm noise}^2 = 74~\times 10^{3}~\rm mK^2$ (as obtained in \citetalias{Mertens2020}). For $\approx$100 nights of observation we scale it down by a factor of 2 for the $f_{\rm sky}$ component and by a factor of 20 for the rest of the components (see Section~\ref{sec:genmock}). The values for $\sigma_{\rm noise}^2$ in each case are listed in the first row for each component.   Note that $\sigma_{\rm noise}^2$ has been scaled to the field of view of the {\tt GRIZZLY} simulations, which is $3.03^{\circ} \times 3.03^{\circ}$. 
}
\begin{tabular}{lllllll}
\hline
Hyperparameter & Input value from & Prior & \multicolumn{2}{c}{$x_{\rm HI}$ fluctuation dominated} & \multicolumn{2}{c}{$T_{\rm S}$ fluctuation dominated} \\
 & \citetalias{Mertens2020} & & 10 nights & 100 nights & 10 nights & 100 nights \\
\hline
$\sigma_{\rm noise}^2$ [ $\times~10^3$ mK$^2$] & & & 74 & 37 & 74 & 37 \\
$\eta_{\rm sky}$ & $+\infty$ & & $+\infty$ & $+\infty$ & $+\infty$ & $+\infty$ \\
$l_{\rm sky}$ & 47.5 & $U (10,100)$ & 43.44$\pm4.02$ & 47.76$\pm1.64$ & 52.00$\pm6.91$ & 45.84$\pm1.67$ \\ 
$\sigma_{\rm sky}^2/\sigma_{\rm noise}^2$ & 611 & -- & 665.17$\pm29.52$ & 579.91$\pm24.11$ & 598.99$\pm27.13$ & 633.43$\pm28.23$ \\
\hline
$\sigma_{\rm noise}^2$ [ $\times~10^3$ mK$^2$] & & & 74 & 3.7 & 74 & 3.7 \\
$\eta_{\rm mix}$ & $3/2$ & & $3/2$ & $3/2$ & $3/2$ & $3/2$ \\
$l_{\rm mix}$ & 2.97 & $U (1,10)$ & 2.99$\pm0.07$ & 2.97$\pm0.08$ & 3.08$\pm0.08$ & 2.90$\pm0.08$ \\ 
$\sigma_{\rm mix}^2/\sigma_{\rm noise}^2$ & 50.4 & -- & 54.80$^{+2.53}_{-2.42}$ & 46.05$^{+2.14}_{-2.04}$ & 55.82$^{+2.64}_{-2.52}$ & 52.12$^{+2.58}_{-2.46}$ \\ 
\hline
$\sigma_{\rm noise}^2$ [ $\times~10^3$ mK$^2$] & & & 74 & 3.7 & 74 & 3.7 \\
$\eta_{\rm ex}$ & $5/2$ & & $5/2$ & $5/2$ & $5/2$ & $5/2$ \\
$l_{\rm ex}$ & 0.26 & $U (0.1,0.8)$ & 0.26$\pm0.003$ & 0.26$\pm0.005$ & 0.27$\pm0.01$ & 0.26$\pm0.01$ \\
$\sigma_{\rm ex}^2/\sigma_{\rm noise}^2$ & 2.18 & -- & 2.15$^{+0.05}_{-0.05}$ & 1.91$^{+0.11}_{-0.10}$ & 2.25$^{+0.07}_{-0.07}$ & 2.32$^{+0.18}_{-0.17}$ \\
\hline
$\sigma_{\rm noise}^2$ [ $\times~10^3$ mK$^2$] & & & 74 & 3.7 & 74 & 3.7 \\
$x_1$ & -- & -- & -0.10$\pm0.99$ & -0.23$\pm1.02$ & -0.41$\pm0.92$ & -0.07$\pm1.00$ \\
$x_2$ & -- & -- & -0.17$\pm0.98$ & 0.42$\pm0.77$ & 0.92$\pm0.59$ & -0.06$\pm0.31$ \\
$\sigma_{\rm 21}^2/\sigma_{\rm noise}^2$ & -- & -- & 0.04$^{+0.21}_{-0.03}$ & 0.17$^{+0.37}_{-0.12}$ & 0.25$^{+0.33}_{-0.14}$ & 1.70$^{+1.07}_{-0.65}$ \\
\hline
\end{tabular}
\label{table:results}
\end{table*}

\subsection{ {\tt CRASH} simulations}\label{sec:MBII_Crash_results}

\begin{figure}
\centering
\includegraphics[width=0.95\columnwidth,keepaspectratio]{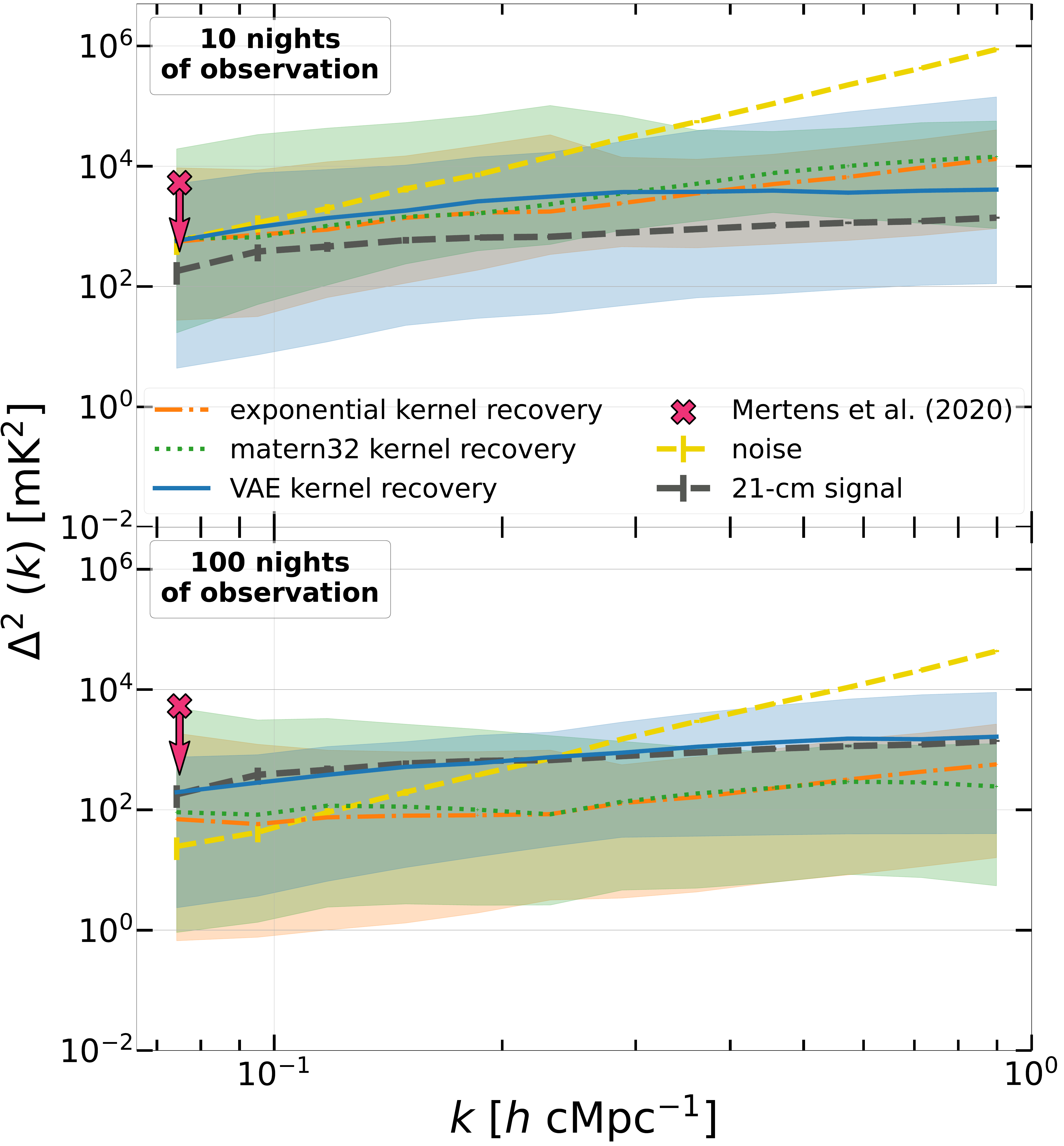}
\caption{As Figure~\ref{fig_outputUTs} for the {\tt CRASH} simulation of reionization.}  
\label{fig_outputFullSim}
\end{figure}

The power spectra resulting from the recovery using the three kernels in the case of the {\tt CRASH} simulation are shown in Figure~\ref{fig_outputFullSim}. As now the input 21-cm signal has a power which lies in between the two {\tt GRIZZLY} models, this translates into an intermediate SNR across $k$-bins. Due to this, we are able to successfully contain the input signal in the 2$\sigma$ ranges of the recovered signals in most $k$-bins for $\approx$10 nights of observation with all kernels, \revised{but given the thermal noise uncertainty power, it is still classified as an upper limit}. However, for $\approx$100 nights, we note that while the VAE kernel does an excellent job of recovering the signal with narrow \revised{2$\sigma$} uncertainty bands, \revised{they still indicate that the recovery is an upper limit. On the other hand,} the Matern class functions based kernels underestimate the signal and do not contain the input 21-cm signal \revised{within 2$\sigma$ uncertainty bands for} some $k$-bins, despite them being significantly larger. \revised{When comparing to the thermal noise uncertainty, we note that they only provide upper limits for the signal in most $k$-bins.} This result is similar to that with the $T_{\rm S}$ fluctuation dominated model with {\tt GRIZZLY}, and thus is due to the same reasons discussed in Section~\ref{sec:varTS_results}.

\subsection{Redshift dependence}\label{sec:redshift}

To evaluate the performance of the VAE kernel at different redshifts, we use the simulations at $z$ = 8.30, 9.16 and 10.11 of both {\tt GRIZZLY} models, and compare the performance of the VAE kernel for noise levels equivalent to $\approx$100 nights of observations. It should be noted that we use the VAE kernels trained for the respective redshifts to avoid making the assumption of the kernels being redshift agnostic. 

\begin{figure}
\centering
\includegraphics[width=0.95\columnwidth,keepaspectratio]{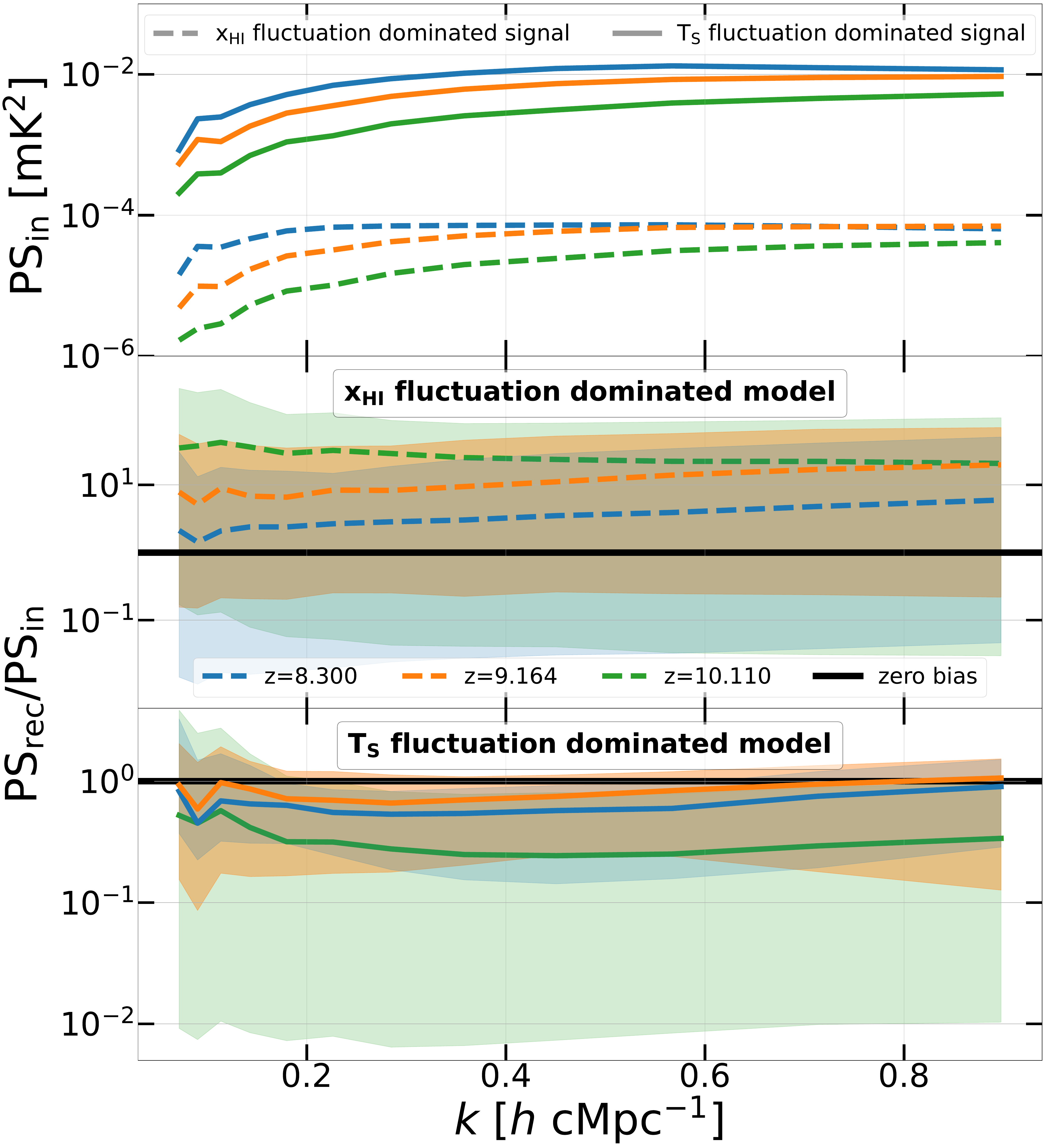}
\caption{\textit{Top panel:} input 21-cm signal for the $x_{\rm HI}$ (dashed lines) and the $T_{\rm S}$  (solid) fluctuation dominated model are shown for $z =$ 8.30 (blue), 9.16 (orange) and 10.11 (green). 
\textit{Middle panel:} recovered 21-cm signal and its associated uncertainty divided by the input 21-cm signal to give the bias and scaled uncertainty (the average values over $k$-bins for these quantities are listed in Table~\ref{table:var_z}) for the $x_{\rm HI}$ fluctuation dominated model. For comparison, the line of zero bias (i.e., recovered signal perfectly matching the input signal) is also shown. 
\textit{Bottom panel:} same as the middle panel, but for the $T_{\rm S}$ fluctuation dominated model.}  
\label{fig_redshiftdep}
\end{figure}

To analyse our recovery technique, we check how \textit{accurately} it recovers the input signal, and how \textit{precise} the results it reports are. For this purpose, we use the quantities defined below, with their values listed in Table~\ref{table:var_z}:
\begin{enumerate}
    \item \textit{Average bias}, $\langle {\rm PS_{\rm rec}}/{\rm PS_{\rm in}}\rangle_k$: this is the average bias given as the ratio between the recovered ($\rm PS_{\rm rec}$) and input 21-cm signal ($\rm PS_{\rm in}$) power spectra, averaged across all $k$-bins. An accurate recovery has a value close to 1, with higher (lower) values indicating an over(under)-estimation of the input 21-cm signal. We note that the deviation of average bias from 1 increases significantly with redshift for the $x_{\rm HI}$ fluctuation dominated model. However, the trend is not so clear for the $T_{\rm S}$ fluctuation dominated signal, as the power and slope of the excess noise component match closely those of the input 21-cm signal at $z = 8.30$ and $k \leq 0.2$ hMpc$^{-1}$, which makes differentiating between them more difficult. However, we still get an average bias of $\approx 0.70$, suggesting only a minor under-estimation, which is primarily observed at the lowest $k$-bins (see Figure~\ref{fig_redshiftdep}).

    \item \textit{Average scaled uncertainty}, $\langle \rm Err_{\rm rec}/\rm PS_{\rm in}\rangle_{k}$: 
    the 2$\sigma$ uncertainty (given by the shaded region in Figures~\ref{fig_outputUTs}, \ref{fig_outputVTs}, \ref{fig_outputFullSim}, and  referred to as $\rm Err_{\rm rec}$ hereafter) on the recovered power spectrum ($\rm PS_{\rm rec}$) gives the precision of recovery. However, \revised{except for cases of extremely poor recovery,} the absolute value of $\rm Err_{\rm rec}$ \revised{generally} depends on the absolute value of the corresponding $\rm PS_{\rm rec}$. \revised{Thus,} to compare different recoveries for a given $\rm PS_{\rm in}$, we need to convert it into a unitless quantity. We first \revised{tried to} do this by calculating the ratio $\rm Err_{\rm rec}/\rm PS_{\rm rec}$ as a measure of the precision of the recovery. However, this ratio contains no information on the accuracy of $\rm PS_{\rm rec}$. Indeed, one could have a precise recovery, i.e. a low $\rm Err_{\rm rec}/\rm PS_{\rm rec}$, but an inaccurate $\rm PS_{\rm rec}$, i.e. a bias which deviates significantly from 1 (see point above). This ratio, then, is not useful, as it does not quantify the overall quality of recovery. To overcome this issue, we use the ratio $\rm Err_{\rm rec}/\rm PS_{\rm in}$ instead. As the magnitude of $\rm Err_{\rm rec}$ depends on that of $\rm PS_{\rm rec}$, it also carries the information of the bias in recovery.
    Further, as we divide by $\rm PS_{\rm in}$, the scaled uncertainty becomes independent of the specific  $\rm PS_{\rm in}$ being recovered. This allows us to use it to compare between cases with different $\rm PS_{\rm in}$, such as at different redshifts (see top panel of Figure~\ref{fig_redshiftdep} where the power spectrum varies due to the time evolution of ionising bubbles) and different physical models (as discussed in Sections~\ref{sec:sims}). We use this generalised comparison, and call it the \textit{scaled} uncertainty for each $k$-bin. Averaging this across all $k$-bins provides a handy quantity to compare the precision of recovery for input 21-cm signals with different physical properties. For example, in Figure~\ref{fig_redshiftdep} we observe that in the $x_{\rm HI}$ fluctuation dominated model, while small-scale variability due to partial reionization is restricted, it still has variability tied to the large-scale distribution of neutral Hydrogen, which increases for lower redshifts. This boosts the power at large scales, corresponding to the low $k$-bins. The same can also be seen in the $T_{\rm S}$ fluctuation dominated model, although its overall power is boosted, as it allows small-scale variability in $\delta T_{\rm b}$ as well. Using the scaled uncertainty, we can compare the precision of recovery across redshift for both cases. The difference in $\langle \rm Err_{\rm rec}/\rm PS_{\rm in}\rangle_{k}$ is quite significant for the $x_{\rm HI}$ fluctuation dominated model, going from $\approx15$ at $z=8.30$ to $\approx80$ at $z=10.11$.
    
    \item \textit{z-score}, $\langle$z-score$\rangle_{k}$: The z-score \citep{Kirch_2008} is a popular quantity to evaluate quality of recovery. In our case, it is defined as $\frac{\rm PS_{\rm rec} - \rm PS_{\rm in}}{\sigma}$ \revised{or $\frac{\rm PS_{\rm rec} - \rm PS_{\rm in}}{\rm Err_{\rm rec}/2}$} at each $k$-bin, and it measures how much the recovered signal deviates from the input 21-cm signal, in units of standard deviation of the recovered signal. The z-score is thus a more explicit method to combine into a single quantity the information provided by the bias along with that of the uncertainty. The only \revised{possible issue is} that $\sigma ( = \rm Err_{\rm rec}/2)$ depends on $\rm PS_{\rm rec}$ and thus their ratio would mask the accuracy of recovery as discussed in the point above (see below for an example case). \revised{Further, we note that we cannot just report an average of z-scores across all $k$-bins, as the distribution of z-scores is not necessarily Gaussian. Indeed, we find that while it is approximately Gaussian for the $x_{\rm HI}$ fluctuation dominated model, this is not the case for the $T_{\rm S}$ fluctuation dominated model. Thus, we report the minimum and maximum z-scores (z-score$_{\rm min}$ and z-score$_{\rm max}$) along with the average ($\langle$z-score$\rangle_{k}$).} \revised{When $\langle$z-score$\rangle_{k}>$($<$)0, its exact value quantifies the extent of over(under)-prediction.} We note that in the $x_{\rm HI}$ fluctuation dominated model, the average z-score worsens with increasing redshift, consistently with the behaviour of the average bias and average scaled uncertainty. This trend is not detected for the $T_{\rm S}$ fluctuation dominated signal, due to the same reasons discussed above for the average bias in (i).
    We also note that at $z=10.11$, $\langle$z-score$\rangle_{k} \approx -2$  and $+1$  in the $T_{\rm S}$ and $x_{\rm HI}$ fluctuation dominated model respectively, naively suggesting a better recovery for the latter. This, though, is not correct, but simply a consequence of the very broad error bars and the inverse proportionality of $\langle$z-score$\rangle_{k}$ with the error. This reasoning exposes the limitation of the z-score. Indeed, by looking at $\langle {\rm PS_{\rm rec}}/{\rm PS_{\rm in}}\rangle_k$ and $\langle \rm Err_{\rm rec}/\rm PS_{\rm in}\rangle_{k}$ for recovery of the $x_{\rm HI}$ fluctuation dominated signal (see Table~\ref{table:var_z}), we note that the deviation from zero bias (obtained when $\langle {\rm PS_{\rm rec}}/{\rm PS_{\rm in}}\rangle_k$ = 1) and zero uncertainty (when $\langle \rm Err_{\rm rec}/\rm PS_{\rm in}\rangle_{k}$ = 0) is significantly higher than for the $T_{\rm S}$ fluctuation dominated model. In fact, these numbers suggest a better quality of recovery in the 
    latter case, which is understandable as the SNR in this model is higher. Thus, while we report the z-score due to its popularity, we recommend using the average bias and scaled uncertainty for evaluating the quality of recovery.
\end{enumerate}

The trends in various quantities discussed above are linked to the physical nature of the 21-cm power spectrum and its redshift evolution. The drop in the SNR with increasing redshift (due to a decrease in signal power as shown in Figure~\ref{fig_redshiftdep} and explained in (ii)), leads to a worsening of the average bias and scaled uncertainty, especially for the $x_{\rm HI}$ fluctuation dominated model. 
As already mentioned, when the excess noise and the input 21-cm signal have similar power and slope (as at $z = 8.30$ for the $T_{\rm S}$ fluctuation dominated case), we observe limitations in the capability of differentiating among the two, but the effects are minor and 
the trends of average bias, scaled uncertainty and z-score for the $x_{\rm HI}$ fluctuation dominated model are also observed in the $T_{\rm S}$ fluctuation dominated model when going from $z = 9.16$ to $z = 10.11$.

Thus, we find that the VAE kernel does not add significant biases, with its recovery and associated uncertainty largely scaling with the physical properties of the 21-cm signal.

\begin{table*}
\centering
\caption{Average bias, $\langle {\rm PS_{\rm rec}}/{\rm PS_{\rm in}}\rangle_k$, average scaled uncertainty, $\langle \rm Err_{\rm rec}/{\rm PS_{\rm in}}\rangle_k$, and average z-score, $\langle {\rm z-score}\rangle_k$, at various redshifts. These values are obtained when the signal recovery is done employing the VAE kernel with the $x_{\rm HI}$ (left) and the $T_{\rm S}$ (right) fluctuation dominated model and $\approx$100 nights of observations. 
}
\begin{tabular}{lllllllllll}
\hline 
$z$ & \multicolumn{5}{c}{$x_{\rm HI}$ fluctuation dominated} & \multicolumn{5}{c}{$T_{\rm S}$ fluctuation dominated} \\
 & $\langle\frac{\rm PS_{\rm rec}}{\rm PS_{\rm in}}\rangle_k$ &$\langle\frac{\rm Err_{\rm rec}}{\rm PS_{\rm in}}\rangle_{k}$ & z-score$_{\rm min}$ & $\langle$z-score$\rangle_{k}$ & z-score$_{\rm max}$ & $\langle\frac{\rm PS_{\rm rec}}{\rm PS_{\rm in}}\rangle_k$ & $\langle\frac{\rm Err_{\rm rec}}{\rm PS_{\rm in}}\rangle_{k}$ & z-score$_{\rm min}$ & $\langle$z-score$\rangle_{k}$ & z-score$_{\rm max}$ \\
\hline
8.30 & 3.90 & 14.88 & \revised{+0.46} & +0.74 & \revised{+0.85} & 0.69 & 0.50 & \revised{-4.00} & -1.36 & \revised{-1.25} \\
9.16 & 8.57 & 21.12 & \revised{+0.73} & +0.82 & \revised{+0.88} & 0.90 & 0.58 & \revised{-0.79} & -0.60 & \revised{+0.23} \\
10.11 &29.29 & 78.22 & \revised{+0.87} & +1.02 & \revised{+1.16} & 0.34 & 0.83 & \revised{-3.50} & -1.96 & \revised{-0.92} \\
\hline
\end{tabular}
\begin{flushleft}
\end{flushleft}
\label{table:var_z}
\end{table*}

\section{Discussion}\label{sec:discuss}

\subsection{Role of the excess noise component}\label{sec:excess}

In \citetalias{Mertens2020} it was noted that the excess noise component was poorly constrained, and thus the combined excess noise and 21-cm signal components were jointly recovered, as separating them was not statistically justifiable. Thus, it is important to understand how well constrained the excess noise component has to be, in order to separate the 21-cm signal from it. 

To explore this, we looked at the $x_{\rm HI}$ fluctuation dominated model of the 21-cm signal at $z = 9.16$. We note that in the $\approx$100 nights case for the chosen excess noise component, the power spectrum recovered with the VAE kernel is slightly overestimated, with the 2$\sigma$ bands on both sides of it spanning $\approx$2 orders of magnitude. We then generated a range of input excess noise components (by varying either the coherence-scale hyperparameter or the variance), and analysed the effects on the recovery of the 21-cm signal power spectrum using the VAE kernel by looking at the average bias, scaled uncertainty and z-score, as defined in Section~\ref{sec:redshift}. The results are reported in Table~\ref{table:var_excessnoise}, 
where f$_{\rm var}$ is the factor
by which the coherence-scale hyperparameter and the variance from the results of \citetalias{Mertens2020} were scaled. We see that, overall, the average bias and scaled uncertainty are reduced when increasing the coherence-scale hyperparameter or decreasing the variance, while no substantial difference is observed in the average z-score. This is possibly because the bias and scaled uncertainty decrease at the same rate, and thus their effects roughly cancel out.

\begin{table}
\centering
\caption{Average bias, $\langle {\rm PS_{\rm rec}}/{\rm PS_{\rm in}}\rangle_k$, average scaled uncertainty, $\langle \rm Err_{\rm rec}/{\rm PS_{\rm in}}\rangle_k$, and average z-score, $\langle {\rm z-score}\rangle_k$, for the $x_{\rm HI}$ fluctuation dominated model with $\approx$100 nights of observations at $z=9.16$.
These values are obtained when the signal recovery is done employing the VAE kernel and multiplying the coherence-scale hyperparameter, $l_{\rm ex}$, or the variance, $\sigma_{\rm ex}^2$, listed in the second column of Table~\ref{table:results} (and as used in M20) by a factor $f_{\rm var}$. } 
\begin{tabular}{llllll}
\hline 
f$_{\rm var}$ & $\langle\frac{\rm PS_{\rm rec}}{\rm PS_{\rm in}}\rangle_k$ &$\langle \frac{\rm Err_{\rm rec}}{\rm PS_{\rm in}}\rangle_{k}$ & z-score$_{\rm min}$ & $\langle$z-score$\rangle_{k}$ & z-score$_{\rm max}$ \\
\hline
\multicolumn{6}{c}{$l_{\rm ex}$} \\
\hline
0.5 &12.88 & 31.37 & \revised{+0.88} & +1.04 & \revised{+1.19} \\
0.75& 9.61 & 26.75 & \revised{+0.77} & +0.89 & \revised{+0.97} \\
1.0 & 8.57 & 21.12 & \revised{+0.73} & +0.82 & \revised{+0.88} \\
1.25& 8.04 & 27.40 & \revised{+0.75} & +0.82 & \revised{+0.89} \\
1.5 & 7.77 & 15.65 & \revised{+0.78} & +0.84 & \revised{+0.90} \\
1.75& 5.64 & 14.76 & \revised{+0.81} & +0.87 & \revised{+0.93} \\
2.0 & 4.55 & 19.38 & \revised{+0.81} & +0.85 & \revised{+0.94} \\
\hline
\multicolumn{6}{c}{$\sigma_{\rm ex}^2$} \\
\hline
1.00 & 8.57 & 21.12 & \revised{+0.73} & +0.82 & \revised{+0.88} \\
0.75 & 6.64 & 32.81 & \revised{+0.75} & +0.80 & \revised{+0.88} \\
0.50 & 4.77 & 16.89 & \revised{+0.66} & +0.71 & \revised{+0.78} \\
0.25 & 3.24 &  9.13 & \revised{+0.58} & +0.62 & \revised{+0.67} \\
\hline
\end{tabular}
\label{table:var_excessnoise}
\end{table}

\subsection{Overall performance of the VAE kernel}\label{sec:VAEkernel_performance}

Usually it is expected that the recovery of the 21-cm signal from an overall dataset with foreground components, noise and signal is not possible if SNR<1. In Section~\ref{sec:uniformTS_results} we have indeed shown that Matern class functions based kernels are unable to contain the input 21-cm signal within their 2$\sigma$ uncertainty bands when SNR<1. However, the VAE kernel is not only able to do so, but also to recover the overall shape of the power spectrum, as seen in the top panel of Figure~\ref{fig_outputUTs}. 

Further, as highlighted by \citet{Kern_2021}, misestimation of the covariance kernel can significantly hamper signal detection \revised{given the currently used normalization and bias correction schemes}. \revised{This means that as} the Matern class functions model the 21-cm signal only approximately, their results can be significantly biased for more complex models of the 21-cm signal, as given e.g. by the $T_{\rm S}$ fluctuation dominated model of {\tt GRIZZLY} and the \revised{21-cm signal model from the} {\tt CRASH} simulations. Indeed, in these cases the Matern class functions based kernels fail to recover the signal also for  noise levels equivalent to $\approx$100 nights of observation, even when $\langle \rm SNR \rangle_k \approx10$. The VAE kernel does not suffer from such a limitation and performs well also when used with an input signal from the {\tt CRASH} simulations. This shows that the VAE covariance kernel is a more robust estimate of the covariance of the 21-cm signal, and can successfully report a detection within 2$\sigma$ uncertainty regardless of the exact physical properties of the observed 21-cm differential brightness temperature power spectrum. Lastly, we note that it performs well across all redshifts analysed here. This reconfirms the robustness of the VAE kernel in constraining the 21-cm signal, with an increase/decrease in uncertainty tied to the $\langle \rm SNR \rangle_k$ of the signal itself.

Overall, the 2$\sigma$ uncertainty bands given by the VAE kernel contain the signal in all cases discussed here. We consider the recovery limit of the VAE kernel in terms of SNR averaged over $k$-bins to be that from $\approx$10 nights of observation in the $x_{\rm HI}$ fluctuation dominated model, i.e. $\langle \rm SNR \rangle_k = 5 \times 10^{-2}$. For values lower than this, we do not expect the VAE kernel based recovery to contain the input signal within its 2$\sigma$ uncertainty bands across $k = [0.05, 1.00] ~h \rm Mpc^{-1}$. We explore one such case  in Appendix~\ref{appendix}, and indeed find that the recovery does not contain the input signal within its 2$\sigma$ uncertainty bands, but instead provides upper limits. Yet it still outperforms the Matern class functions based kernels in recovering the shape of the power spectrum, spread of uncertainty on recovery, as well as the reported upper limits.

\subsection{Limitations}\label{sec:limitations}

\revised{In this work, we reduce the possibility of biases in the EoR covariance kernel by incorporating a more physically motivated covariance. We showcase the robustness of the generated VAE covariance kernel by testing it against not just mock 21-cm signals obtained with {\tt GRIZZLY}, but with signals generated using very different frameworks as shown in Section~\ref{sec:MBII_Crash_results} and in Appendix~\ref{appendix}. This has also been demonstrated in \citet{Mertens_2023}.

However, biases are still possible, especially if the true signal, and thus its covariance, is fundamentally different from what we obtain with our simulation codes. One way to further minimise this bias is to use mock data obtained from different codes to train the covariance kernel, which we plan to do in further upgrades of our pipeline. In the future, we will also investigate methods to reduce the dependence on the prior by using different normalization and bias correction schemes as suggested by \cite{Kern_2021}.}

\section{Summary}
The LOFAR Epoch of Reionization (EoR) KSP team strives for a successful detection of the 21-cm signal from the EoR at $z \approx 7-11$. Past theoretical models indicated that at least 1000 hours of observation would be necessary to lead to a successful detection \citep{Mertens2018}, while \citet{Mertens2020} provided upper limits using 141 hours ($\approx$10 nights) of observation. In this respect, an optimal choice of the covariance kernel for the 21-cm signal component is crucial. Indeed, as shown in \citet{Kern_2021}, \revised{given the currently used normalization and bias correction scheme,} a mismatch between the adopted and the actual covariance kernel of the 21-cm signal can induce a  significant signal loss, which can in turn lead to incorrect astrophysical interpretations from any ``successful'' detection. 

To improve the choice of the 21-cm signal covariance kernel, \citet{Mertens_2023} introduce a Machine Learning method which employs a Variational Auto-Encoder (VAE) based algorithm. As the training done using VAE is not limited by the form of the specific function (as e.g. in the case of Matern class functions), nor by the kernels of the training datasets (as in the case of a simple Auto-Encoder), it allows to reproduce the covariance kernel of the 21-cm signal with a greater flexibility. This is showcased in terms of the robustness of the VAE based kernel's performance in comparison to previously used kernels based on Matern class functions.

We show that the result on using the VAE kernel is able to contain the input 21-cm signal within its 2$\sigma$ uncertainty band in all cases explored  where $\langle \rm SNR \rangle_k \gtrsim 5 \times 10^{-2}$. It is also usually better than the results from Matern class functions based covariance kernels in recovering the overall shape of the power spectrum of the signal. A key result in this paper is that Matern class functions based kernels are unable to recover the 21-cm signal for the $T_{\rm S}$ fluctuation dominated model even for $\approx$100 nights of observation, for which $\langle \rm SNR \rangle_k \approx10$, while a recovery with the VAE kernel is successful. A similar result is obtained also with a 21-cm signal generated using the {\tt CRASH} simulations, thus clearly indicating that the Matern class functions based kernels do not correctly match the covariance of more complex models of the signal. Thus, this analysis suggests that using the VAE kernel can mitigate to a significant extent the issues highlighted by \citet{Kern_2021} \revised{given no change to the normalization and bias correction schemes}. 

Further, we show that the behaviour of the VAE kernel is consistent across all redshifts of interest, with changes in its performance strongly correlating with the neutral hydrogen distribution, which changes the strength of the resultant power spectrum, and thus the $\langle \rm SNR \rangle_k$. This suggests that the VAE based kernel can be used for any choice of redshift without additional correction factors, making the algorithm developed here directly applicable to LOFAR data at $z \approx 8~\rm to~10$, whose results can then be compared with results from telescopes like HERA.

We also explore the effects that the properties of the excess noise component identified in \citetalias{Mertens2020} have on the recovery of the 21-cm signal. As expected, we find that having a higher coherence scale or a lower variance for the components leads to better recovery.

In companion papers we will apply the VAE kernel to the $\approx$10 nights of LOFAR data used in \citetalias{Mertens2020}, and  explore the range of theoretical models which are consistent with the upper limits provided by the VAE kernel, as done in \citet{Ghara2020}. Applying the VAE kernel to observations much longer than $\approx$10 nights requires a significant improvement in the modelling of the intrinsic sky component, which would eventually be limited by the confusion noise due to the angular resolution of LOFAR. Further improvements, such as noise mitigation, can be implemented by choosing data from nights with better ionospheric conditions and lesser contribution from RFI flagging. All these aspects are currently being explored by the LOFAR EoR KSP team.

\section*{Acknowledgements}
AA thanks the EoR research group at MPA, as well as Hitesh Kishore Das and Abinaya Swaruba Rajamuthukumar for helpful discussions. 
FM acknowledges support of the PSL Fellowship. RG acknowledges support by the Israel Science Foundation (grant no. 255/18). LVEK and SM acknowledge the financial support from the European Research Council (ERC) under the European Union's Horizon 2020 research and innovation programme (Grant agreement No. 884760, "CoDEX"). The post-doctoral contract of IH was funded by Sorbonne Université in the framework of the Initiative Physique des Infinis (IDEX SUPER). QM is supported by the National Natural Science Foundation of China (Grant No. 12263002). GM acknowledges support by the Swedish Research Council grant 2020-04691. Nordita is supported in part by NordForsk.
\revised{Lastly, we would like to thank the referee for their helpful comments that helped improve this paper significantly.}

\section*{Data Availability}
The VAE kernels generated for $z = 8.30, 9.16, 10.11$ \revised{shall be made public on publication. The {\tt ps\_eor} package used to run GPR and ML-GPR can be found at \url{https://gitlab.com/flomertens/ps_eor}.
The} underlying training, testing and validation datasets generated using {\tt GRIZZLY}, and the mock 21-cm signal data generated using {\tt GRIZZLY} and {\tt CRASH} can be shared upon reasonable request. 


\bibliographystyle{mnras}
\bibliography{mlgpr}

\begin{thebibliography}{}
\makeatletter
\relax
\def\mn@urlcharsother{\let\do\@makeother \do\$\do\&\do\#\do\^\do\_\do\%\do\~}
\def\mn@doi{\begingroup\mn@urlcharsother \@ifnextchar [ {\mn@doi@}
  {\mn@doi@[]}}
\def\mn@doi@[#1]#2{\def\@tempa{#1}\ifx\@tempa\@empty \href
  {http://dx.doi.org/#2} {doi:#2}\else \href {http://dx.doi.org/#2} {#1}\fi
  \endgroup}
\def\mn@eprint#1#2{\mn@eprint@#1:#2::\@nil}
\def\mn@eprint@arXiv#1{\href {http://arxiv.org/abs/#1} {{\tt arXiv:#1}}}
\def\mn@eprint@dblp#1{\href {http://dblp.uni-trier.de/rec/bibtex/#1.xml}
  {dblp:#1}}
\def\mn@eprint@#1:#2:#3:#4\@nil{\def\@tempa {#1}\def\@tempb {#2}\def\@tempc
  {#3}\ifx \@tempc \@empty \let \@tempc \@tempb \let \@tempb \@tempa \fi \ifx
  \@tempb \@empty \def\@tempb {arXiv}\fi \@ifundefined
  {mn@eprint@\@tempb}{\@tempb:\@tempc}{\expandafter \expandafter \csname
  mn@eprint@\@tempb\endcsname \expandafter{\@tempc}}}

\bibitem[\protect\citeauthoryear{{Aigrain} \& {Foreman-Mackey}}{{Aigrain} \&
  {Foreman-Mackey}}{2023}]{Aigrain_2023}
{Aigrain} S.,  {Foreman-Mackey} D.,  2023, \mn@doi [\araa]
  {10.1146/annurev-astro-052920-103508}, \href
  {https://ui.adsabs.harvard.edu/abs/2023ARA&A..61..329A} {61, 329}

\bibitem[\protect\citeauthoryear{{Becker} et~al.,}{{Becker}
  et~al.}{2001}]{Becker_2001}
{Becker} R.~H.,  et~al., 2001, \mn@doi [\aj] {10.1086/324231}, \href
  {https://ui.adsabs.harvard.edu/abs/2001AJ....122.2850B} {122, 2850}

\bibitem[\protect\citeauthoryear{{Becker}, {Bolton}, {Madau}, {Pettini},
  {Ryan-Weber}  \& {Venemans}}{{Becker} et~al.}{2015}]{Becker_2015}
{Becker} G.~D.,  {Bolton} J.~S.,  {Madau} P.,  {Pettini} M.,  {Ryan-Weber}
  E.~V.,   {Venemans} B.~P.,  2015, \mn@doi [\mnras] {10.1093/mnras/stu2646},
  \href {https://ui.adsabs.harvard.edu/abs/2015MNRAS.447.3402B} {447, 3402}

\bibitem[\protect\citeauthoryear{{Bosman} et~al.,}{{Bosman}
  et~al.}{2022}]{Bosman_2022}
{Bosman} S. E.~I.,  et~al., 2022, \mn@doi [\mnras] {10.1093/mnras/stac1046},
  \href {https://ui.adsabs.harvard.edu/abs/2022MNRAS.514...55B} {514, 55}

\bibitem[\protect\citeauthoryear{{Ciardi} \& {Ferrara}}{{Ciardi} \&
  {Ferrara}}{2005}]{Ciardi_2005}
{Ciardi} B.,  {Ferrara} A.,  2005, \mn@doi [\ssr] {10.1007/s11214-005-3592-0},
  \href {https://ui.adsabs.harvard.edu/abs/2005SSRv..116..625C} {116, 625}

\bibitem[\protect\citeauthoryear{{Ciardi} \& {Madau}}{{Ciardi} \&
  {Madau}}{2003}]{Ciardi_2003}
{Ciardi} B.,  {Madau} P.,  2003, \mn@doi [\apj] {10.1086/377634}, \href
  {https://ui.adsabs.harvard.edu/abs/2003ApJ...596....1C} {596, 1}

\bibitem[\protect\citeauthoryear{{Ciardi}, {Ferrara}, {Marri}  \&
  {Raimondo}}{{Ciardi} et~al.}{2001}]{Ciardi_2001}
{Ciardi} B.,  {Ferrara} A.,  {Marri} S.,   {Raimondo} G.,  2001, \mn@doi
  [\mnras] {10.1046/j.1365-8711.2001.04316.x}, \href
  {https://ui.adsabs.harvard.edu/abs/2001MNRAS.324..381C} {324, 381}

\bibitem[\protect\citeauthoryear{{Cl{\'e}ment} et~al.,}{{Cl{\'e}ment}
  et~al.}{2012}]{Clement_2012}
{Cl{\'e}ment} B.,  et~al., 2012, \mn@doi [\aap] {10.1051/0004-6361/201117312},
  \href {https://ui.adsabs.harvard.edu/abs/2012A&A...538A..66C} {538, A66}

\bibitem[\protect\citeauthoryear{{Davies} et~al.,}{{Davies}
  et~al.}{2018}]{Davies_2018}
{Davies} F.~B.,  et~al., 2018, \mn@doi [\apj] {10.3847/1538-4357/aad6dc}, \href
  {https://ui.adsabs.harvard.edu/abs/2018ApJ...864..142D} {864, 142}

\bibitem[\protect\citeauthoryear{{Eide}, {Graziani}, {Ciardi}, {Feng},
  {Kakiichi}  \& {Di Matteo}}{{Eide} et~al.}{2018}]{Eide_2018}
{Eide} M.~B.,  {Graziani} L.,  {Ciardi} B.,  {Feng} Y.,  {Kakiichi} K.,   {Di
  Matteo} T.,  2018, \mn@doi [\mnras] {10.1093/mnras/sty272}, \href
  {https://ui.adsabs.harvard.edu/abs/2018MNRAS.476.1174E} {476, 1174}

\bibitem[\protect\citeauthoryear{{Eide}, {Ciardi}, {Graziani}, {Busch}, {Feng}
  \& {Di Matteo}}{{Eide} et~al.}{2020}]{Eide_2020}
{Eide} M.~B.,  {Ciardi} B.,  {Graziani} L.,  {Busch} P.,  {Feng} Y.,   {Di
  Matteo} T.,  2020, \mn@doi [\mnras] {10.1093/mnras/staa2774}, \href
  {https://ui.adsabs.harvard.edu/abs/2020MNRAS.498.6083E} {498, 6083}

\bibitem[\protect\citeauthoryear{{Fan} et~al.,}{{Fan} et~al.}{2006}]{Fan_2006}
{Fan} X.,  et~al., 2006, \mn@doi [\aj] {10.1086/500296}, \href
  {https://ui.adsabs.harvard.edu/abs/2006AJ....131.1203F} {131, 1203}

\bibitem[\protect\citeauthoryear{{Foreman-Mackey}, {Hogg}, {Lang}  \&
  {Goodman}}{{Foreman-Mackey} et~al.}{2013}]{Foreman_Mackey_2013}
{Foreman-Mackey} D.,  {Hogg} D.~W.,  {Lang} D.,   {Goodman} J.,  2013, \mn@doi
  [\pasp] {10.1086/670067}, \href
  {https://ui.adsabs.harvard.edu/abs/2013PASP..125..306F} {125, 306}

\bibitem[\protect\citeauthoryear{Furlanetto}{Furlanetto}{2016}]{Furlanetto_2016}
Furlanetto S.~R.,  2016, The 21-cm Line as a Probe of Reionization.
Springer International Publishing, Cham, pp 247--280,
  \mn@doi{10.1007/978-3-319-21957-8_9}, \url
  {https://doi.org/10.1007/978-3-319-21957-8_9}

\bibitem[\protect\citeauthoryear{{Furlanetto}, {Oh}  \&
  {Pierpaoli}}{{Furlanetto} et~al.}{2006}]{Furlanetto_2006eq}
{Furlanetto} S.~R.,  {Oh} S.~P.,   {Pierpaoli} E.,  2006, \mn@doi [\prd]
  {10.1103/PhysRevD.74.103502}, \href
  {https://ui.adsabs.harvard.edu/abs/2006PhRvD..74j3502F} {74, 103502}

\bibitem[\protect\citeauthoryear{{Gehlot} et~al.,}{{Gehlot}
  et~al.}{2019}]{Gehlot_2019}
{Gehlot} B.~K.,  et~al., 2019, \mn@doi [\mnras] {10.1093/mnras/stz1937}, \href
  {https://ui.adsabs.harvard.edu/abs/2019MNRAS.488.4271G} {488, 4271}

\bibitem[\protect\citeauthoryear{{Ghara}, {Choudhury}  \& {Datta}}{{Ghara}
  et~al.}{2015}]{Ghara_2015}
{Ghara} R.,  {Choudhury} T.~R.,   {Datta} K.~K.,  2015, \mn@doi [\mnras]
  {10.1093/mnras/stu2512}, \href
  {https://ui.adsabs.harvard.edu/abs/2015MNRAS.447.1806G} {447, 1806}

\bibitem[\protect\citeauthoryear{{Ghara}, {Mellema}, {Giri}, {Choudhury},
  {Datta}  \& {Majumdar}}{{Ghara} et~al.}{2018}]{Ghara_2018}
{Ghara} R.,  {Mellema} G.,  {Giri} S.~K.,  {Choudhury} T.~R.,  {Datta} K.~K.,
  {Majumdar} S.,  2018, \mn@doi [\mnras] {10.1093/mnras/sty314}, \href
  {https://ui.adsabs.harvard.edu/abs/2018MNRAS.476.1741G} {476, 1741}

\bibitem[\protect\citeauthoryear{{Ghara} et~al.,}{{Ghara}
  et~al.}{2020}]{Ghara2020}
{Ghara} R.,  et~al., 2020, \mn@doi [\mnras] {10.1093/mnras/staa487}, \href
  {https://ui.adsabs.harvard.edu/abs/2020MNRAS.493.4728G} {493, 4728}

\bibitem[\protect\citeauthoryear{{Glatzle}, {Ciardi}  \& {Graziani}}{{Glatzle}
  et~al.}{2019}]{Glatzle_2019}
{Glatzle} M.,  {Ciardi} B.,   {Graziani} L.,  2019, \mn@doi [\mnras]
  {10.1093/mnras/sty2514}, \href
  {https://ui.adsabs.harvard.edu/abs/2019MNRAS.482..321G} {482, 321}

\bibitem[\protect\citeauthoryear{Goodfellow, Bengio  \& Courville}{Goodfellow
  et~al.}{2016}]{Goodfellow_2016}
Goodfellow I.,  Bengio Y.,   Courville A.,  2016, Deep Learning.
MIT Press

\bibitem[\protect\citeauthoryear{{Graziani}, {Maselli}  \& {Ciardi}}{{Graziani}
  et~al.}{2013}]{Graziani_2013}
{Graziani} L.,  {Maselli} A.,   {Ciardi} B.,  2013, \mn@doi [\mnras]
  {10.1093/mnras/stt206}, \href
  {https://ui.adsabs.harvard.edu/abs/2013MNRAS.431..722G} {431, 722}

\bibitem[\protect\citeauthoryear{{Graziani}, {Ciardi}  \& {Glatzle}}{{Graziani}
  et~al.}{2018}]{Graziani_2018}
{Graziani} L.,  {Ciardi} B.,   {Glatzle} M.,  2018, \mn@doi [\mnras]
  {10.1093/mnras/sty1367}, \href
  {https://ui.adsabs.harvard.edu/abs/2018MNRAS.479.4320G} {479, 4320}

\bibitem[\protect\citeauthoryear{{Greig} \& {Mesinger}}{{Greig} \&
  {Mesinger}}{2015}]{Greig2015}
{Greig} B.,  {Mesinger} A.,  2015, \mn@doi [\mnras] {10.1093/mnras/stv571},
  \href {https://ui.adsabs.harvard.edu/abs/2015MNRAS.449.4246G} {449, 4246}

\bibitem[\protect\citeauthoryear{{Greig}, {Mesinger}, {Haiman}  \&
  {Simcoe}}{{Greig} et~al.}{2017}]{Greig_2017}
{Greig} B.,  {Mesinger} A.,  {Haiman} Z.,   {Simcoe} R.~A.,  2017, \mn@doi
  [\mnras] {10.1093/mnras/stw3351}, \href
  {https://ui.adsabs.harvard.edu/abs/2017MNRAS.466.4239G} {466, 4239}

\bibitem[\protect\citeauthoryear{{Hogan} \& {Rees}}{{Hogan} \&
  {Rees}}{1979}]{Hogan_1979}
{Hogan} C.~J.,  {Rees} M.~J.,  1979, \mn@doi [\mnras]
  {10.1093/mnras/188.4.791}, \href
  {https://ui.adsabs.harvard.edu/abs/1979MNRAS.188..791H} {188, 791}

\bibitem[\protect\citeauthoryear{{Hothi} et~al.,}{{Hothi}
  et~al.}{2021}]{Hothi_2021}
{Hothi} I.,  et~al., 2021, \mn@doi [\mnras] {10.1093/mnras/staa3446}, \href
  {https://ui.adsabs.harvard.edu/abs/2021MNRAS.500.2264H} {500, 2264}

\bibitem[\protect\citeauthoryear{{Kern} \& {Liu}}{{Kern} \&
  {Liu}}{2021}]{Kern_2021}
{Kern} N.~S.,  {Liu} A.,  2021, \mn@doi [\mnras] {10.1093/mnras/staa3736},
  \href {https://ui.adsabs.harvard.edu/abs/2021MNRAS.501.1463K} {501, 1463}

\bibitem[\protect\citeauthoryear{{Kerrigan} et~al.,}{{Kerrigan}
  et~al.}{2018}]{Kerrigan_2018}
{Kerrigan} J.~R.,  et~al., 2018, \mn@doi [\apj] {10.3847/1538-4357/aad8bb},
  \href {https://ui.adsabs.harvard.edu/abs/2018ApJ...864..131K} {864, 131}

\bibitem[\protect\citeauthoryear{{Khandai}, {Di Matteo}, {Croft}, {Wilkins},
  {Feng}, {Tucker}, {DeGraf}  \& {Liu}}{{Khandai} et~al.}{2015}]{Khandai_2015}
{Khandai} N.,  {Di Matteo} T.,  {Croft} R.,  {Wilkins} S.,  {Feng} Y.,
  {Tucker} E.,  {DeGraf} C.,   {Liu} M.-S.,  2015, \mn@doi [\mnras]
  {10.1093/mnras/stv627}, \href
  {https://ui.adsabs.harvard.edu/abs/2015MNRAS.450.1349K} {450, 1349}

\bibitem[\protect\citeauthoryear{{Kingma} \& {Welling}}{{Kingma} \&
  {Welling}}{2013}]{Kingma_2013}
{Kingma} D.~P.,  {Welling} M.,  2013, \mn@doi [arXiv e-prints]
  {10.48550/arXiv.1312.6114}, \href
  {https://ui.adsabs.harvard.edu/abs/2013arXiv1312.6114K} {p. arXiv:1312.6114}

\bibitem[\protect\citeauthoryear{{Kingma} \& {Welling}}{{Kingma} \&
  {Welling}}{2019}]{Kingma_2019}
{Kingma} D.~P.,  {Welling} M.,  2019, \mn@doi [arXiv e-prints]
  {10.48550/arXiv.1906.02691}, \href
  {https://ui.adsabs.harvard.edu/abs/2019arXiv190602691K} {p. arXiv:1906.02691}

\bibitem[\protect\citeauthoryear{Kirch}{Kirch}{2008}]{Kirch_2008}
Kirch W.,  ed. 2008, z-Score.
Springer Netherlands, Dordrecht, pp 1484--1484,
  \mn@doi{10.1007/978-1-4020-5614-7_3826}, \url
  {https://doi.org/10.1007/978-1-4020-5614-7_3826}

\bibitem[\protect\citeauthoryear{Kullback \& Leibler}{Kullback \&
  Leibler}{1951}]{Kullback_1951}
Kullback S.,  Leibler R.~A.,  1951, The annals of mathematical statistics, 22,
  79

\bibitem[\protect\citeauthoryear{{Liu} \& {Shaw}}{{Liu} \&
  {Shaw}}{2020}]{Liu_2020}
{Liu} A.,  {Shaw} J.~R.,  2020, \mn@doi [\pasp] {10.1088/1538-3873/ab5bfd},
  \href {https://ui.adsabs.harvard.edu/abs/2020PASP..132f2001L} {132, 062001}

\bibitem[\protect\citeauthoryear{{Ma}, {Ciardi}, {Eide}, {Busch}, {Mao}  \&
  {Zhi}}{{Ma} et~al.}{2021}]{Ma_2021}
{Ma} Q.-B.,  {Ciardi} B.,  {Eide} M.~B.,  {Busch} P.,  {Mao} Y.,   {Zhi} Q.-J.,
   2021, \mn@doi [\apj] {10.3847/1538-4357/abefd5}, \href
  {https://ui.adsabs.harvard.edu/abs/2021ApJ...912..143M} {912, 143}

\bibitem[\protect\citeauthoryear{{Madau}, {Meiksin}  \& {Rees}}{{Madau}
  et~al.}{1997}]{Madau_1997}
{Madau} P.,  {Meiksin} A.,   {Rees} M.~J.,  1997, \mn@doi [\apj]
  {10.1086/303549}, \href
  {https://ui.adsabs.harvard.edu/abs/1997ApJ...475..429M} {475, 429}

\bibitem[\protect\citeauthoryear{{Maselli}, {Ciardi}  \& {Kanekar}}{{Maselli}
  et~al.}{2009}]{Maselli_2009}
{Maselli} A.,  {Ciardi} B.,   {Kanekar} A.,  2009, \mn@doi [\mnras]
  {10.1111/j.1365-2966.2008.14197.x}, \href
  {https://ui.adsabs.harvard.edu/abs/2009MNRAS.393..171M} {393, 171}

\bibitem[\protect\citeauthoryear{{Mertens}, {Ghosh}  \& {Koopmans}}{{Mertens}
  et~al.}{2018}]{Mertens2018}
{Mertens} F.~G.,  {Ghosh} A.,   {Koopmans} L.~V.~E.,  2018, \mn@doi [\mnras]
  {10.1093/mnras/sty1207}, \href
  {https://ui.adsabs.harvard.edu/abs/2018MNRAS.478.3640M} {478, 3640}

\bibitem[\protect\citeauthoryear{{Mertens} et~al.,}{{Mertens}
  et~al.}{2020}]{Mertens2020}
{Mertens} F.~G.,  et~al., 2020, \mn@doi [\mnras] {10.1093/mnras/staa327}, \href
  {https://ui.adsabs.harvard.edu/abs/2020MNRAS.493.1662M} {493, 1662}

\bibitem[\protect\citeauthoryear{{Mertens}, {Bobin}  \& {Carucci}}{{Mertens}
  et~al.}{2023}]{Mertens_2023}
{Mertens} F.~G.,  {Bobin} J.,   {Carucci} I.~P.,  2023, \mn@doi [arXiv
  e-prints] {10.48550/arXiv.2307.13545}, \href
  {https://ui.adsabs.harvard.edu/abs/2023arXiv230713545M} {p. arXiv:2307.13545}

\bibitem[\protect\citeauthoryear{{Mesinger} \& {Furlanetto}}{{Mesinger} \&
  {Furlanetto}}{2007}]{Mesinger2007}
{Mesinger} A.,  {Furlanetto} S.,  2007, \mn@doi [\apj] {10.1086/521806}, \href
  {https://ui.adsabs.harvard.edu/abs/2007ApJ...669..663M} {669, 663}

\bibitem[\protect\citeauthoryear{{Morales} \& {Wyithe}}{{Morales} \&
  {Wyithe}}{2010}]{Morales_2010}
{Morales} M.~F.,  {Wyithe} J. S.~B.,  2010, \mn@doi [\araa]
  {10.1146/annurev-astro-081309-130936}, \href
  {https://ui.adsabs.harvard.edu/abs/2010ARA&A..48..127M} {48, 127}

\bibitem[\protect\citeauthoryear{{Morales}, {Beardsley}, {Pober}, {Barry},
  {Hazelton}, {Jacobs}  \& {Sullivan}}{{Morales} et~al.}{2019}]{Morales_2019}
{Morales} M.~F.,  {Beardsley} A.,  {Pober} J.,  {Barry} N.,  {Hazelton} B.,
  {Jacobs} D.,   {Sullivan} I.,  2019, \mn@doi [\mnras]
  {10.1093/mnras/sty2844}, \href
  {https://ui.adsabs.harvard.edu/abs/2019MNRAS.483.2207M} {483, 2207}

\bibitem[\protect\citeauthoryear{Mortlock}{Mortlock}{2016}]{Mortlock_2016}
Mortlock D.,  2016, Quasars as Probes of Cosmological Reionization.
Springer International Publishing, Cham, pp 187--226,
  \mn@doi{10.1007/978-3-319-21957-8_7}, \url
  {https://doi.org/10.1007/978-3-319-21957-8_7}

\bibitem[\protect\citeauthoryear{Pinheiro~Cinelli, Ara{\'u}jo~Marins, Barros~da
  Silva  \& Lima~Netto}{Pinheiro~Cinelli et~al.}{2021}]{Cinelli_2021}
Pinheiro~Cinelli L.,  Ara{\'u}jo~Marins M.,  Barros~da Silva E.~A.,
  Lima~Netto S.,  2021, Variational Autoencoder.
Springer International Publishing, Cham, pp 111--149,
  \mn@doi{10.1007/978-3-030-70679-1_5}, \url
  {https://doi.org/10.1007/978-3-030-70679-1_5}

\bibitem[\protect\citeauthoryear{{Planck Collaboration} et~al.,}{{Planck
  Collaboration} et~al.}{2016}]{Planck_2016}
{Planck Collaboration} et~al., 2016, \mn@doi [\aap]
  {10.1051/0004-6361/201628897}, \href
  {https://ui.adsabs.harvard.edu/abs/2016A&A...596A.108P} {596, A108}

\bibitem[\protect\citeauthoryear{{Pober} et~al.,}{{Pober}
  et~al.}{2014}]{Pober_2014}
{Pober} J.~C.,  et~al., 2014, \mn@doi [\apj] {10.1088/0004-637X/782/2/66},
  \href {https://ui.adsabs.harvard.edu/abs/2014ApJ...782...66P} {782, 66}

\bibitem[\protect\citeauthoryear{{Pritchard} \& {Loeb}}{{Pritchard} \&
  {Loeb}}{2012}]{Pritchard_2012}
{Pritchard} J.~R.,  {Loeb} A.,  2012, \mn@doi [Reports on Progress in Physics]
  {10.1088/0034-4885/75/8/086901}, \href
  {https://ui.adsabs.harvard.edu/abs/2012RPPh...75h6901P} {75, 086901}

\bibitem[\protect\citeauthoryear{Rasmussen \& Williams}{Rasmussen \&
  Williams}{2006}]{RasmussenBook}
Rasmussen C.~E.,  Williams C. K.~I.,  2006, Gaussian processes for machine
  learning.
Adaptive computation and machine learning, MIT Press

\bibitem[\protect\citeauthoryear{{Schenker} et~al.,}{{Schenker}
  et~al.}{2013}]{Schenker_2013}
{Schenker} M.~A.,  et~al., 2013, \mn@doi [\apj] {10.1088/0004-637X/768/2/196},
  \href {https://ui.adsabs.harvard.edu/abs/2013ApJ...768..196S} {768, 196}

\bibitem[\protect\citeauthoryear{{Shaver}, {Windhorst}, {Madau}  \& {de
  Bruyn}}{{Shaver} et~al.}{1999}]{Shaver_1999}
{Shaver} P.~A.,  {Windhorst} R.~A.,  {Madau} P.,   {de Bruyn} A.~G.,  1999,
  \aap, \href {https://ui.adsabs.harvard.edu/abs/1999A&A...345..380S} {345,
  380}

\bibitem[\protect\citeauthoryear{Stein}{Stein}{1999}]{Stein1999}
Stein M.~L.,  1999, Interpolation of spatial data: some theory for kriging.
Springer Science \& Business Media

\bibitem[\protect\citeauthoryear{{The HERA Collaboration} et~al.,}{{The HERA
  Collaboration} et~al.}{2022}]{HERA_2022}
{The HERA Collaboration} et~al., 2022, arXiv e-prints, \href
  {https://ui.adsabs.harvard.edu/abs/2022arXiv221004912T} {p. arXiv:2210.04912}

\bibitem[\protect\citeauthoryear{{Tozzi}, {Madau}, {Meiksin}  \&
  {Rees}}{{Tozzi} et~al.}{2000}]{Tozzi_2000}
{Tozzi} P.,  {Madau} P.,  {Meiksin} A.,   {Rees} M.~J.,  2000, \mn@doi [\apj]
  {10.1086/308196}, \href
  {https://ui.adsabs.harvard.edu/abs/2000ApJ...528..597T} {528, 597}

\bibitem[\protect\citeauthoryear{Zaroubi}{Zaroubi}{2013}]{Zaroubi_2013}
Zaroubi S.,  2013, The Epoch of Reionization.
Springer Berlin Heidelberg, Berlin, Heidelberg, pp 45--101,
  \mn@doi{10.1007/978-3-642-32362-1_2}, \url
  {https://doi.org/10.1007/978-3-642-32362-1_2}

\makeatother
\end{thebibliography}


\appendix

\section{Investigating a low SNR case}\label{appendix}

We use the semi-numerical code {\tt 21cmFAST} \citep{Mesinger2007, Greig2015} to generate a mock 21-cm signal at $z=9.1$ within a box of length 400 cMpc. We adopt the values $\zeta = 30\%$, $R_{\rm mfp} = 15~\rm cMpc$ and $T_{\rm vir}^{\rm feed} = 5 \times 10^4$ K (see \citealt{Greig2015} for more details). 

\begin{figure}
\centering
\includegraphics[width=0.95\columnwidth,keepaspectratio]{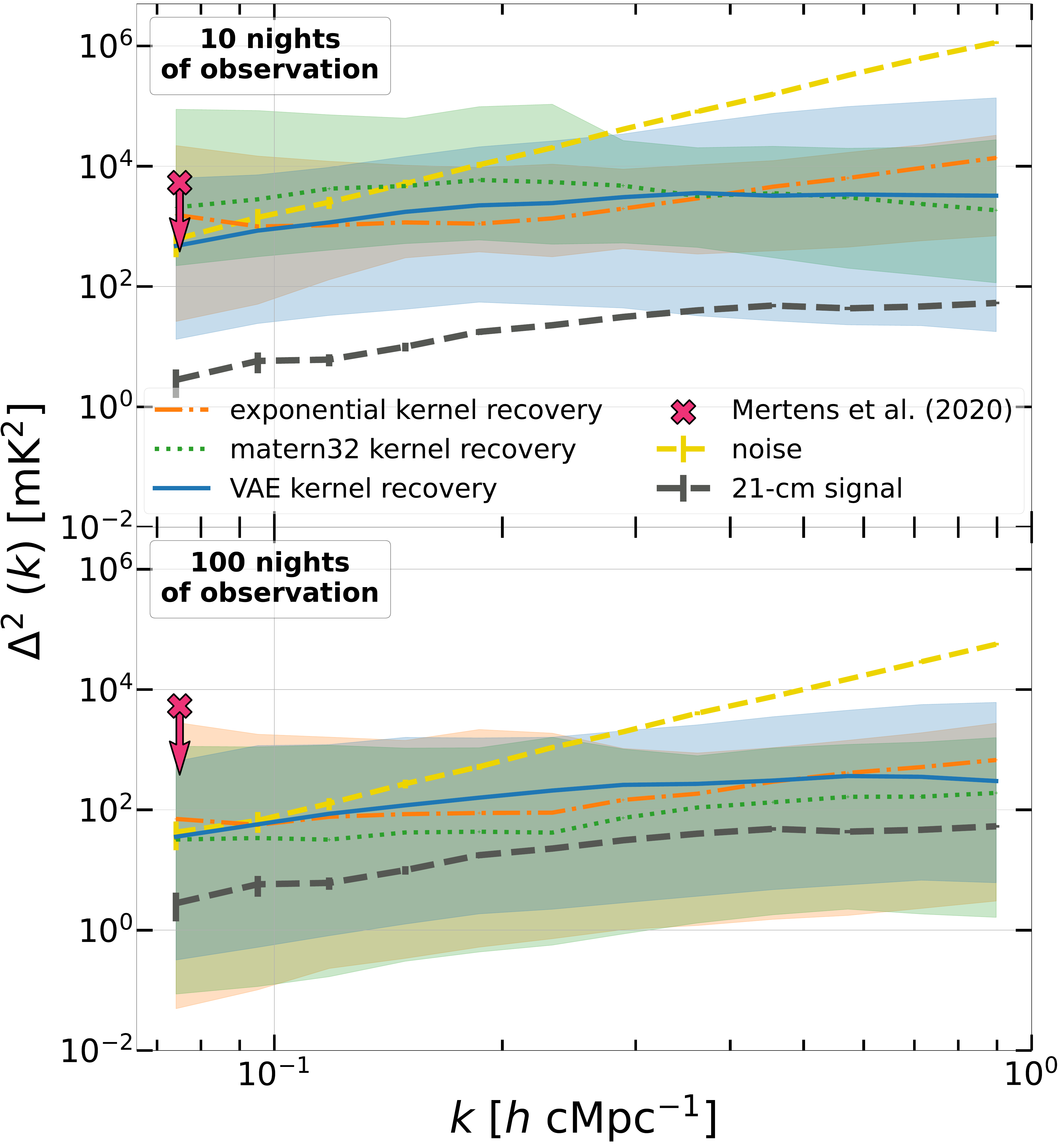}
\caption{As Figure~\ref{fig_outputUTs} for {\tt 21cmFAST} simulations of reionization.}  
\label{fig_output21cmFAST}
\end{figure}

The power spectra recovered using the three kernels are shown in Figure~\ref{fig_output21cmFAST}. As in this model the input 21-cm signal is much weaker than in those discussed in the main text, for $\approx 10$ nights of observation ($\langle \rm SNR \rangle_{\sl k} \approx 1.6 \times 10^{-2}$) the input signal is never contained within the 2$\sigma$ uncertainty bounds of the recovered signal with the exception of the highest $k$-bins, where the SNR is higher. \revised{When comparing with the thermal noise uncertainty, we find that all three kernels manage to provide upper limits, with the VAE kernel also recovering the overall shape. A detection is not possible because of the low SNR, with the uncertainty on the thermal noise being almost two orders of magnitude larger than the input 21-cm signal.} For $\approx$100 nights ($\langle \rm SNR \rangle_{\sl k} \approx 0.3$), the signal is within the 2$\sigma$ uncertainty bounds for all kernels, \revised{while still providing upper limits and not a detection because the SNR is still low despite the increased integration time. The VAE kernel has} the tightest constraints on the uncertainty. While the VAE kernel overestimates the input signal, it does an excellent job in recovering its overall shape.

\bsp	
\label{lastpage}
\end{document}